\documentclass{aa}  

\usepackage{graphicx}
\usepackage{txfonts}
\newcommand{\re}{$R_e$}
\newcommand{\Lsig}{$L-\sigma$}

\newcommand{\Lsigb}{$L=L'_0\sigma^{\beta}$}
\newcommand{\Lsigbtempo}{$L=L'_{0}(t) \sigma^{\beta(t)}$}                   %%%nuovo

\newcommand{\MRa}{$R_e$-$M^*$}
\newcommand{\Rsigma}{$R_e - \sigma$}

\newcommand{\Ie}{$I_e$}

\newcommand{\IeRe}{$I_e - R_e$}
\newcommand{\IeSig}{$I_e - \sigma$}

\newcommand{\FPR}{$\log(\sigma)-\log(I_e)-\log(R_e)$}

\newcommand{\ie}{{i.e.}}
\newcommand{\eg}{{e.g.}}

\newcommand{\kms}{km\, sec$^{-1}$}

\begin{document}

   \title{The scaling relations of galaxies back in time: \\
   the road toward virialization}

   %\subtitle{I. Overviewing the $\kappa$-mechanism}

   \author{M. D'Onofrio\fnmsep\thanks{Corresponding author: Mauro D'Onofrio}
          \inst{1}
          \and
          C. Chiosi\inst{1}
          }

   \institute{Department of Physics and Astronomy, University of Padua,
              Vicolo Osservatorio 3, I-35122 Padova (Italy)\\
              \email{mauro.donofrio@unipd.it} \\
              \email{cesare.chiosi@unipd.it}
         %\and
          %   University of Alexandria, Department of Geography, ...\\
           %  \email{c.ptolemy@hipparch.uheaven.space}
            % \thanks{The university of heaven temporarily does not
             %        accept e-mails}
             }

   \date{Received January 2023; accepted June 2023}

% \abstract{}{}{}{}{} 
% 5 {} token are mandatory
 
  \abstract
  % context heading (optional)
  % {} leave it empty if necessary 
   {The structural scaling relations (SSRs) of galaxies, \ie\  the observed correlations between effective radius, effective surface intensity and velocity dispersion, are important tools for understanding how evolution proceeds.}
  % aims heading (mandatory)
   {In this paper we aim to demonstrate that the evolution of the SSRs back in time is governed by the combination of the virial theorem (VT) and the relation \Lsigbtempo, where the parameters $\beta$ and $L'_0$ vary with time and from galaxy to galaxy.}
  % methods heading (mandatory)
   {Using the WINGS database for the galaxies at redshift $z=0$ and the Illustris-1 and Illustris-TNG databases of artificial galaxies, for the galaxies up to redshift $z=4$, we analyse the SSRs back in time and, by means of simple algebraic expressions for $L'_0$ and $\beta$ (functions of time and other physical quantities), we derive the expected paths followed by galaxies in the various SSRs toward the distributions observed at $z=0$.}
  % results heading (mandatory)
   {The distribution of galaxies in the SSRs is ultimately related to the evolution in luminosity and velocity dispersion that are empirically mirrored  by the \Lsigbtempo\ law. Furthermore, the $\beta$ parameter works as a thermometer of the virialization of a galaxy. This parameter can assume either positive or negative values, and its absolute value attains  high values when the galaxy is close to the virial condition, while it tends to zero when the galaxy is far from it.}
  % conclusions heading (optional), leave it empty if necessary 
   {As the SSRs change with time, the method we are proposing allows us to decipher the temporal evolution of galaxies. } 


   \maketitle
%
%-------------------------------------------------------------------

\section{Introduction}
\label{sec:1}

The structural scaling relations (SSRs) of galaxies, \ie\ the mutual correlations between the main measured structural parameters, such as the effective radius \re, the effective surface intensity \Ie, the total stellar mass $M_s$, the  luminosity $L$, and the central velocity dispersion $\sigma$, have been recognized long ago as important tools for understanding the evolution of these stellar systems and for deriving fundamental cosmological information 
\cite[see \eg,][]{FaberJackson1976,Kormendy1977,Dressleretal1987,DjorgovskiDavis1987}.
In particular, the  SSRs of early-type galaxies (ETGs), that are much easier to obtain, have been used in the past as distance indicators for measuring the Hubble constant \citep[see \eg,][]{Dressler1987}, for testing the expansion of the Universe \citep[see \eg,][]{Pahreetal1996}, for mapping the velocity fields of galaxies \citep[see \eg,][]{DresslerFaber1990,Courteauetal1993}, and for measuring the variation of the mass-to-light ratio across time \cite[see \eg,][]{PrugnielSimien1996,vanDokkumFranx1996,Busarelloetal1998,Franxetal2008}.

Among the various SSRs, the Fundamental Plane (FP) relation for the ETGs \citep{DjorgovskiDavis1987,Dressleretal1987} $\log R_e = a \log\sigma + b \log I_e + c$, is probably the most studied one of the last 30 years. The tilt of this relation with respect to the prediction of the virial theorem  (VT) has been the subject of many studies
\citep[see among many others,][]{Faberetal1987,Ciotti1991,Jorgensenetal1996,Cappellarietal2006,Donofrioetal2006,Boltonetal2007}, invoking different physical mechanisms at work. We remember for example: i) the systematic change of the stellar mass-to-light ratio ($M_s/L$) \citep[see \eg,][]{Faberetal1987,vanDokkumFranx1996,Cappellarietal2006,vanDokkumvanderMarel2007, Holdenetal2010,deGraafetal2021}; ii) the structural and dynamical non-homology of ETGs \citep[see \eg,][]{PrugnielSimien1997,Busarelloetal1998,Trujilloetal2004,Donofrioetal2008}; iii) the dark matter content and distribution \citep[see \eg,][]{Ciottietal1996,Borrielloetal2003,Tortoraetal2009, Taranuetal2015,deGraafetal2021};  iv) the star formation history (SFH) and initial mass function (IMF) \citep[see \eg,][]{RenziniCiotti1993,Chiosietal1998,Chiosi_Carraro_2002,Allansonetal2009}; v) the effects of environment \citep[see \eg,][]{Luceyetal1991,deCarvalhoDjorgovski1992,Bernardietal2003, Donofrioetal2008, LaBarberaetal2010, Ibarra-MedelLopez-Cruz2011, Samiretal2016}; vi) the effects of dissipation-less mergers \citep{Nipotietal2003}; vii) the gas dissipation \citep{robertsonetal06}; viii) the non regular sequence of mergers with progressively decreasing mass ratios \citep{Novak2008}; ix) the multiple dry mergers of spiral galaxies \citep{Taranuetal2015}. 

A similar long list can be compiled for the small intrinsic scatter of the FP ($\approx0.05$ dex in the V-band), where among the claimed possible physical causes, we have: 1) the variation in the formation epoch; 2) the dark matter content; 3) the metallicity or age trends; 4) the variations of the mass-to-light ratio $M/L$ \citep[see e.g.,][]{Faberetal1987, Gregg1992, GuzmanLuceyBower1993, Forbesetal1998, Bernardietal2003, Redaetal2005, Cappellarietal2006, Boltonetal2008, Augeretal2010, Magoulasetal2012}, etc.. 

Despite all these efforts, it is still unclear today why the FP is so tight and uniform when seen edge-on, while in its projections (\ie, in the \IeRe, the \IeSig\ and \Rsigma\ planes) the distribution of galaxies presents well defined structures, where regions with large clumps of objects and big scatter are observed together with regions where no galaxies are present (the so called Zone of Exclusions, ZOE), and where clear non linear  distributions are well visible. The mutual dependence of the SSRs, the peculiar shape of the observed distributions and the link among the various  FP projections have never found a single and robust explanation in which the tilt and the scatter of the FP are understood.

The same difficulties are encountered when we consider one particular projection of the FP: the Faber-Jackson relation (FJ) \citep{FaberJackson1976}, \ie\ the correlation observed between the total luminosity $L$ and the central velocity dispersion $\sigma$. Even in this case the observed trend is not that predicted by the VT. In addition to this, it has been shown that the \Lsig\ relation is not consistent with the distribution observed in the \IeRe\ plane, in the sense that it is not possible to transform one space into the other (and viceversa) adopting the observed classical correlations \citep[see \eg,][]{DonofrioChiosi2021}.

In other words, the underlying questions behind the nature of the SSRs of galaxies are: can we find a single explanation for the tilt of the FP and the shapes of the distributions observed in its projections? 
Is it possible to reconcile the FJ relation and the \IeRe\ plane?
Is it possible to account for the mutual relationship among the various projections of the FP? How are these planes linked to each other? How the SSRs change going back in time? Why the FP is so tight?

A new perspective to simultaneously explain the tilt of the FP and the observed distributions of galaxies in the FP projection planes has been advanced by \citet{Donofrioetal2017}. The novelty of their approach is based on the assumption that the luminosity of galaxies follows a relation of the form: 
\begin{equation}
 L(t)  = L'_0(t) \sigma(t)^{\beta(t)}.
 \label{eq1}
\end{equation}
where $t$ is the time, $\sigma$ the velocity dispersion, and the proportionality coefficient $L'_0$ and the exponent $\beta$ are all function of time and, even more importantly, they can vary from galaxy to galaxy.

This empirical relation is formally equivalent to the FJ relation for ETGs, but has a profoundly different physical meaning. In this relation $\beta$ and $L'_0$ are free time-dependent parameters that can vary considerably from galaxy to galaxy, according to the mass assembly history and the evolution of the stellar content of  each object. The new relation mirrors the effects of the evolutionary history of a galaxy on its luminosity and stellar velocity dispersion, parameters that can both vary across time because galaxies evolve, merge, and interact. 

In previous papers on this subject we called attention on some of the advantages offered by the joint use of the VT and the \Lsigbtempo\ law \citep{Donofrioetal2017,Donofrioetal2019,Donofrioetal2020,DonofrioChiosi2021,Donofrio_Chiosi_2022,Donofrio_Chiosi_2023}. Accepting the idea of a variable $\beta$ parameter, taking either positive or negative values, yields  a simple explanation of the shifts of galaxies along the SSRs. Furthermore it allows us to understand the physical reasons for the observed  distributions of galaxies in the various projection planes. This approach seems to be the correct one because it is able to simultaneously account for: i) the tilt of the FP, ii) the existence of the ZoE,  and iii) the shifts of galaxies in the FP projections that are closely connected with the variations of $\sigma$ and $L$ through the $\beta$ parameter. 

In the present study we take advantage of  what we learned from joining the VT and the law \Lsigbtempo\ to analyze how galaxies move along the SSRs at high redshift. To this aim, since current observational data at high redshifts are not enough for our aims, we adopt the data of the Illustris-1 \citep{Vogelsberger_2014a,Vogelsberger_2014b}  and the Illustris-TNG \citep{Springeletal2018,Nelsonetal2018,Pillepichetal2018a}  simulations from $z=0$ up to $z=4$  and look at the possible changes in the properties of galaxies suggested by the simulations.

The paper is organized as follows: in Sec. \ref{sec:2} we briefly describe the samples of galaxies (both real and simulated) we have used in our work, we present the basic SSRs at $z=0$ and we explain why one can trust in the simulated data at higher redshift. In Sec. \ref{sec:3} we summarize the basic equations of the problem and in Sec. \ref{sec:4} we show how the SSRs change with redshift and how the $\beta$ parameter is able to account of the observed distributions at each epoch. In Sec. \ref{sec:5} we discuss the $\beta$ parameter as a thermometer of the virialization condition. In Sec. \ref{sec:6} we discuss the history of mass assembly for a few test galaxies and investigate how $\beta$ changes as function of time and history of mass assembly. In Sec. \ref{sec:7} we present our conclusions. Finally, in Appendix \ref{appendix_A} we present a toy model of dry and wet mergers to estimate the variation of a galaxy luminosity as consequence of merger and companion star formation. 
  
For the sake of internal consistency with the previous studies of this series, in our calculations with the Illustris-1 database we adopt the same values of the $\Lambda$-CDM cosmology used by  \citep{Vogelsberger_2014a,Vogelsberger_2014b}:
$\Omega_m$ = 0.2726, $\Omega_{\Lambda}$= 0.7274, $\Omega_b$ = 0.0456, $\sigma_8$ = 0.809, $n_s$ = 0.963, $H_0 = 70.4\,km\, s^{-1}\, Mpc^{-1}$. Slightly different cosmological  parameters are used by for the Illustris-TNG simulations: $\Omega_m$ = 0.3089, $\Omega_{\Lambda}$= 0.6911, $\Omega_b$ = 0.0486, $\sigma_8$ = 0.816, $n_s$ = 0.967, $H_0 = 67.74\,km\, s^{-1}\, Mpc^{-1}$ \citep{Springeletal2018,Nelsonetal2018,Pillepichetal2018a} . Since the systematic differences  in $M_s$, $R_e$, $L$, $I_e$, and $\sigma$ are either small or nearly irrelevant to the aims of this study, no re-scaling of the data is applied. 

\section{Observational data and model galaxies}
\label{sec:2}

The observational data used here are the same adopted in our previous works on this subject \citep[see,][]{Donofrio_Chiosi_2022,Donofrio_Chiosi_2023}. The data at redshift $z\sim 0$ have been extracted from the WINGS and Omega-WINGS databases
\citep{Fasanoetal2006,Varela2009,Cava2009,Valentinuzzi2009,Moretti2014,Donofrio2014,Gullieuszik2015,Morettietal2017,Cariddietal2018,Bivianoetal2017}. 

{
The samples used here have not the same dimension in each plot because the spectroscopic database is only a sub-sample of the whole optical photometric sample (containing $\sim 32700$ galaxies). For this reason in some of our plots we can appreciate the distribution of the whole photometric sample, while in others only the subsamples with available measured stellar velocity dispersion or available stellar masses are visible.

%%%%%%%%%%%%Figure 1
   \begin{figure*}
   \centering
   \includegraphics[scale=0.70]{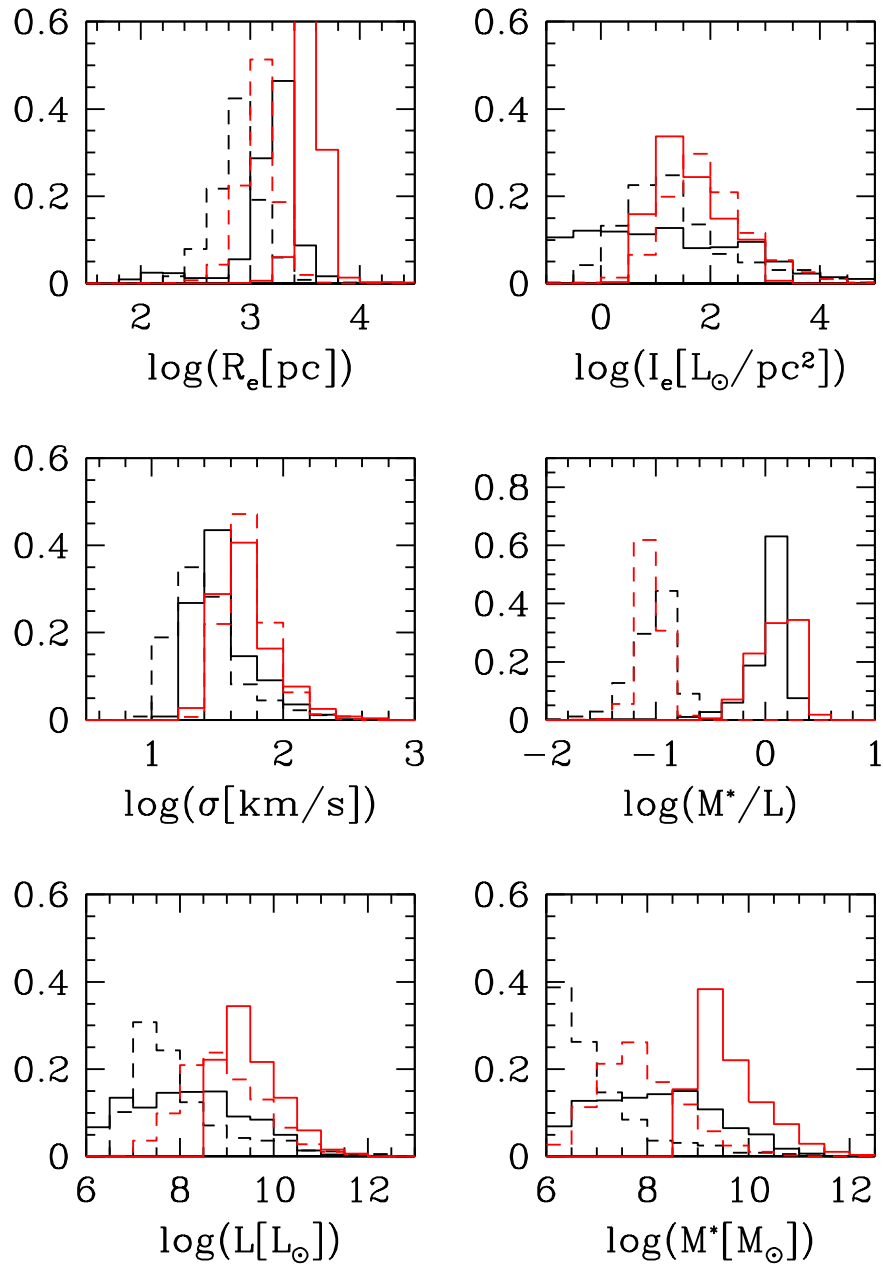}
   \caption{Comparison between the data of Illustris-1 and Illustris-TNG100. The black lines mark the TNG data, while the red ones the Illustris-1 data. The solid line refers to the data at $z=0$, while the dashed lines to the data at $z=4$. From top left to bottom right we have: the effective radius \re\ (enclosing half the total luminosity or half the total stellar mass (for TNG); the effective surface brightness \Ie; the central star velocity dispersion $\sigma$; the stellar mass-to-light ratio ($M^*/L$); the total luminosity (in solar units) and the total stellar mass (in solar units).}
              \label{fig:1}
    \end{figure*}
    
The subsample with measured stellar masses $M_s$ contains approximately 1200 galaxies. The masses were estimated by \cite{Fritzetal2007} by means of the spectral synthesis analysis. This provided the measurements of the stellar masses and of the star formation rate (SFR) at different cosmic epochs (among many others quantities).

The cross-match between the spectroscopic and photometric samples gives here only 480 ETGs with available masses and velocity dispersions. The sample span a magnitude range from $M_V\sim-16$ to $M_V\sim-23$ mag, a central velocity dispersion range from $\sigma\sim50$ to $\sigma\sim300$ \kms\ and masses from $10^{8.5}$ to $10^{12}$ solar masses\footnote{The measured parameters for the real galaxies are always shown in our plots with the black dots. For the reason just explained in each plot, containing real observations, the number of galaxies is not always the same.}. 

The morphological types of the galaxies were measured with the software MORPHOT for the whole photometric dataset. The final morphological type $T$ is quite robust, coming from the combination of different approaches \citep[see][for more details]{Fasanoetal2012}. 
}

The error on the measured parameters is $\simeq20\%$. These are not shown in our plots, because they are much lower than the observed range of variation of the structural parameters in the SSRs. The small size of the errors does not  affect the whole distribution of galaxies. Furthermore, no quantitative analysis has been made here, such as fits of data or statistical evaluations. 

The sample of real data at $z\sim0$ is used only to demonstrate that the simulated galaxies quite well reproduce the SSRs of the local objects and therefore there are good reasons to trust in the simulation when we look at the behavior of the SSRs at much higher redshift.

The analysis of the SSRs at high redshift is unfortunately still difficult for galaxies above $z\sim1.0$, because the observational surveys at these redshifts contain only few and sparse data. Some empirical evidences however exists for a varying tilt of the FP with redshift \citep[see \eg,][]{DiSeregoetal2005, Beifiorietal2017, Lindsayetal2017,deGraaffetal2021}.

Given such difficulties we decided to perform our analysis of the SSRs at high redshift using the database of artificial galaxies provided by the Illustris-1 and Illustris-TNG  simulations. The hydrodynamic simulations, like the Illustris databases, are today the best models available to compare theory with observations, despite the fact that several problems still bias their results.

The first set of artificial galaxies, named Illustris-1 appeared on 2014 \citep[see e.g.][]{Vogelsberger_2014b,Genel_etal_2014,Nelsonetal2015}. Later on,
a number of works demonstrated that Illustris-1 suffer from a number of problems: it yields an unrealistic population of ETGs with no correct colours, it lacks morphological information, the sizes of the less massive galaxies are too big, and the star formation rates are not always comparable with observations \citep[see \eg,][]{Snyderetal2015, Bottrelletal2017,Nelsonetal2018, Rodriguez-Gomezetal2019,Huertas-Companyetal2019,Donofrio_Chiosi_2023}. In addition to this, there is the claim in the literature that Illustris-1 does not produce a realistic red sequence of galaxies due to insufficient quenching of the star formation with too few red galaxies \citep{Snyderetal2015, Bottrelletal2017,Nelsonetal2018,Rodriguez-Gomezetal2019}, while Illustris-TNG produces a much better result \citep{Nelsonetal2018, Rodriguez-Gomezetal2019}. There is also the problem of the insufficient number of red galaxies with respect to the observed population of ETGs. For what concern the internal structure of the Illustris-1 galaxies, \citet{Bottrelletal2017} measured the S\'ersic index, the axis ratio and the radii, and found that too few bulge-dominated objects are produced in tension with observations. In contrast, the Illustris-TNG galaxies have much better internal structural parameters \citep{Rodriguez-Gomezetal2019}. 
For this reason Illustris-1 was superseded in 2018 by Illustris-TNG  \citep{Springeletal2018,Nelsonetal2018,Pillepichetal2018b}. 

{In this work we considered only the subsample named Illustris-TNG-100, which is briefly referred to below as Illustris-TNG.  This sample has approximately the same volume and resolution of Illustris-1 and it used the same initial condition (updated for the different cosmology) adopted by Illustris-1.}

Among the many tabulated quantities provided for the galaxies of Illustris-1, we worked in particular with the V-band photometry, the mass and half-mass radii of the stellar particles (i.e., integrated stellar populations), for the most massive clusters, for which Cartesian comoving coordinates (x', y', z') are available.
We have analyzed in our previous papers the projected light and mass profiles using the z'=0 plane as a reference plane. Starting from the V magnitudes and positions of the stellar particles, we computed the effective radius \re, the radial surface brightness profile in units of $r/R_e$, the best-fit S\'ersic index, and the line-of-sight velocity dispersion.
The values of \re\ were calculated considering only the star particles inside the friend-of-friends (FoFs) of galaxies and the galaxies inside the FoFs of clusters. We have set z'=0 to project the coordinates of the stellar particles inside galaxies so that the velocity dispersion is calculated along the z'-axis. The sample does not contain galaxies with masses lower than $10^9$ solar masses at $z=0$ because for these objects it was impossible to derive \re. The total stellar mass has been used here.
{The data-set for each value of the redshift extracted from the  Illustris-1 simulation and used here contains $\sim 2400$ galaxies of all morphological types.} A full description of this data-set was given in \cite{Cariddietal2018} and \cite{Donofrioetal2020}.

%%%%%%%%%%%%Figure 2
   \begin{figure*}
   \centering
   \includegraphics[scale=0.35]{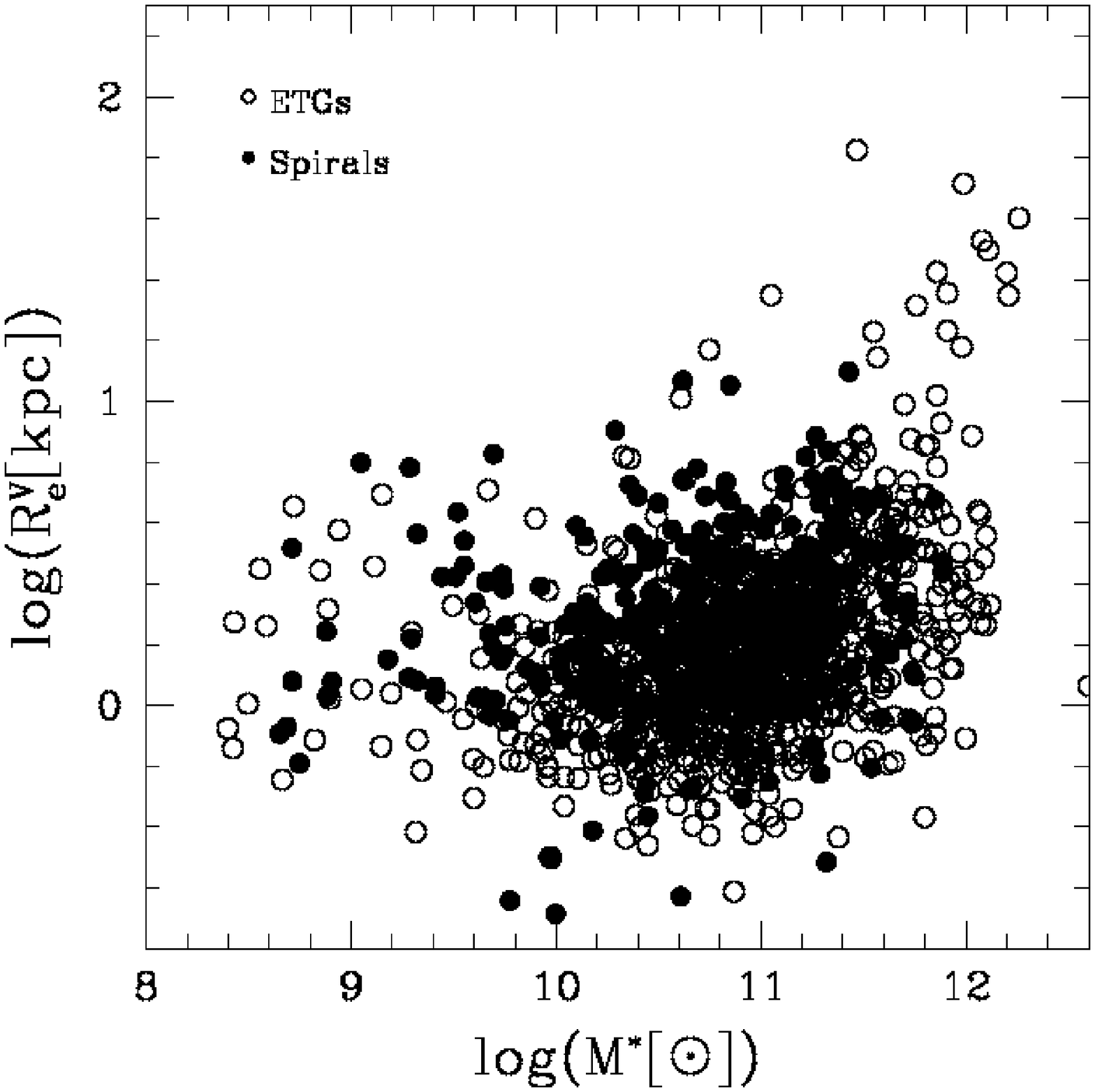}
      \includegraphics[scale=0.4]{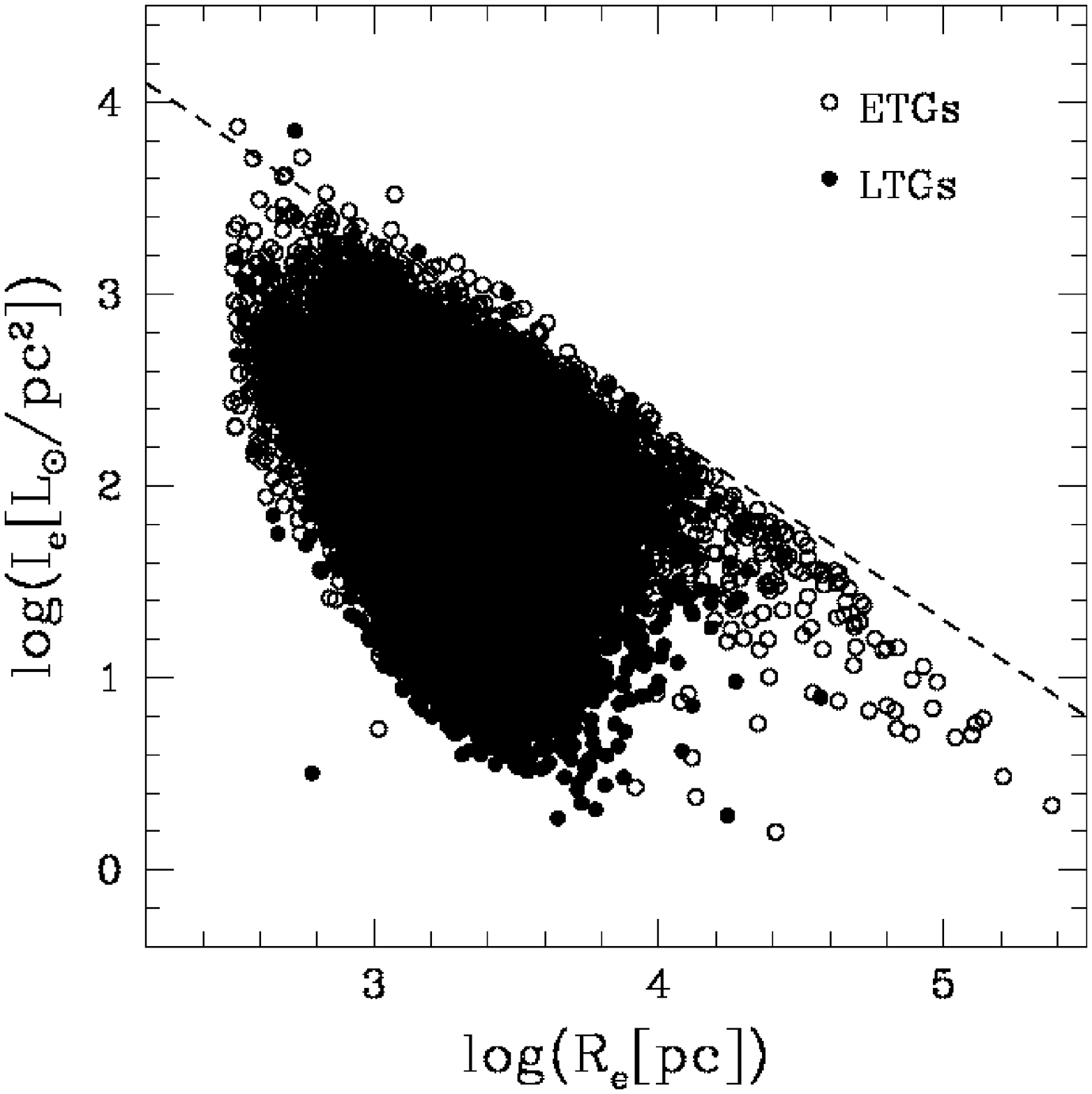}
   \caption{The \MRa\ (left panel) and the \IeRe\ (right panel) planes for the WINGS galaxy sample. Late type galaxies (LTGs) (black dots) and ETGs (open circles) share the same distributions in these SSRs. The effective radius \re\ is given in kpc in the \MRa\ plane and in pc in the \IeRe\ plane. Masses are given in solar units.}
    \label{fig:2}
    \end{figure*}

{From the TNG-100 dataset we selected the first 1000 objects, ordered with decreasing stellar masses, coming out from the online Search Galaxy/Subhalo Catalog\footnote{See https://www.tng-project.org/data/}. In this case we used the half-mass stellar radius instead of the effective radius \re. This radius is not so different from the effective radius and its use does not change in any way the conclusions reached here. The data have been extracted at redshift $z=4$, $z=3$, $z=2$, $z=1$ and $z=0$ in order to be consistent with those used for Illustris-1.}

The choice of using both Illustris-1 and Illustris-TNG has the following reasons: i) we want to be consistent with our previous works on this subject; ii)  the differences in $M_s$, $R_e$, $I_e$, $L$, and $\sigma$  of Illustris-1 and Illustris-TNG do not bias significantly the results on the values of the $\beta$ and $L'_0$ parameters of the \Lsigb\ law \citep[see][]{Donofrio_Chiosi_2023}; iii)  the two data samples are in some way complementary, since Illustris-TNG has better measurements of the half-mass radii of less massive galaxies, while Illustris-1 seems much rich in massive objects; iv) the two simulations agree in the physical parameters of the massive objects.

The detailed analysis of the differences between Illustris-1 and Illustris-TNG data has not been addressed here because there are already several studies on this subject \citep[see \eg,][]{Pillepichetal2018a,Pillepichetal2018b,Rodriguez-Gomezetal2019, Huertas-Companyetal2019}. One of the issues of major tension between the two suites of models concerns the radii of the low mass galaxies (roughly of $M_s \leq 5\,10^{10}\,M_\odot$ where the Illustris-TNG radii are about a factor of two smaller that those of Illustris-1  while above it they are nearly equal \citep{Pillepichetal2018a,Pillepichetal2018b,Rodriguez-Gomezetal2019,Huertas-Companyetal2019}.

{Figure \ref{fig:1} shows the distributions of the Illustris-1 (red lines) and Illustris-TNG100 (black lines) data for several parameters used here at two redshift epochs: $z=0$ (solid lines) and $z=4$ (dashed lines).
From the figure we see that the effective radii of Illustris-1 are systematically a bit larger than those of TNG100. Another significant difference is found in the distribution of the total luminosity and total stellar masses. As already said, the Illustris-1 sample does not contain objects with masses lower than $10^9$ solar masses at $z=0$. It follows that the distribution of masses and luminosities appears different for the TNG sample: it is much smooth and flatter than that of Illustris-1 which seems to be peaked at 9 dex approximately for the objects at $z=0$. The range covered by luminosities and masses however is quite similar.
The other parameters appear more or less superposed. We will see later that such differences do not compromise the analysis done here as well as the main conclusions.
The intrinsic problems of the simulations are of little relevance for our analysis because: i) we do not make use of the color of galaxies and of the SFRs; ii) we have demonstrated \citep[see,][]{Donofrio_Chiosi_2023} that the two samples of Illustris-1 and Illustris-TNG produce very similar distributions of the $\beta$ and $L'_0$ parameters of the \Lsigb\ law; iii) we will show here that the SSRs at high redshift of the two samples are very similar; iv) the point mass view of the galaxies adopted here secures that our analysis is not too much affected by the problems of the simulations.
}
    
For both Illustris-1 and Illustris-TNG we did not extract information on the morphology of the galaxies. For this reason in our comparison ETGs and late-type galaxies (LTGs) are mixed in our plots. This choice originates from the observation that the SSRs of ETGs and LTGs are almost identical. This is clearly seen in Figure \ref{fig:2} showing the \MRa\ (left panel) and the \IeRe\ (right panel) for the ETGs (open circles) and LTGs (filled black circles).
The two distributions are very well superposed in both diagrams. The only exception is that very large \re\ are observed only for the most massive ETGs.
This is only partially in agreement with Fig. 11 of \cite{Huertas-Companyetal2019}, showing that ETGs and LTGs follow quite similar trends, but with small systematic differences for the two morphological types. The data of the WINGS database do not suggest any significant difference in the SSRs of LTGs and ETGs. We believe that the effective radii \re\ measured in their work are affected by a systematic bias due to the method used to derive \re. While for the WINGS galaxies the effective radius was measured as the circle enclosing half the total luminosity, in the Huertas-Company work it was used the semi-major axis of the best fitting S\'ersic model. This choice can likely introduce a systematic effect due to the inclination of the galaxies and the intrinsic shape of the light profiles.  In any case the inclusion of LTGs is a potential source of bias.

We remark in addition that the completeness of the samples is not critical for the conclusions drawn here. In  fact we do not attempt any statistical analysis of the data nor we fit any distribution to derive correlations. The data are only used to  qualitatively  show how the distribution of galaxies in the various planes can change with the different cosmic epochs and how the \Lsigb\ law and the $\beta$ parameter can  at least qualitatively account for the variations expected/observed across time.
The kind of analysis carried out here is indeed somehow independent of the level of precision reached by the models of the different sources, because we are mainly interested in  presenting the method for deciphering the information encrypted in the observational data of the SSRs. The only hypothesis made here is that we can trust the results of simulations at high  redshifts. {This hypothesis is based on the fact that the simulations are able to reproduce some features of the distributions seen in the FP projections at redshifts $z\sim 0$ and the tilt of the FP at $z\sim1$ (see below). The artificial galaxies match quite well the observations, reproducing the position of the brightest cluster galaxies and the existence of the Zone of Exclusion (ZoE). All this  makes us confident that the simulations produce galaxies with luminosities, stellar masses and effective radii not too far from those of real galaxies.}

%%%%%%%%%%%%Figure 3
   \begin{figure*}
   \centering
   \includegraphics[scale=0.35]{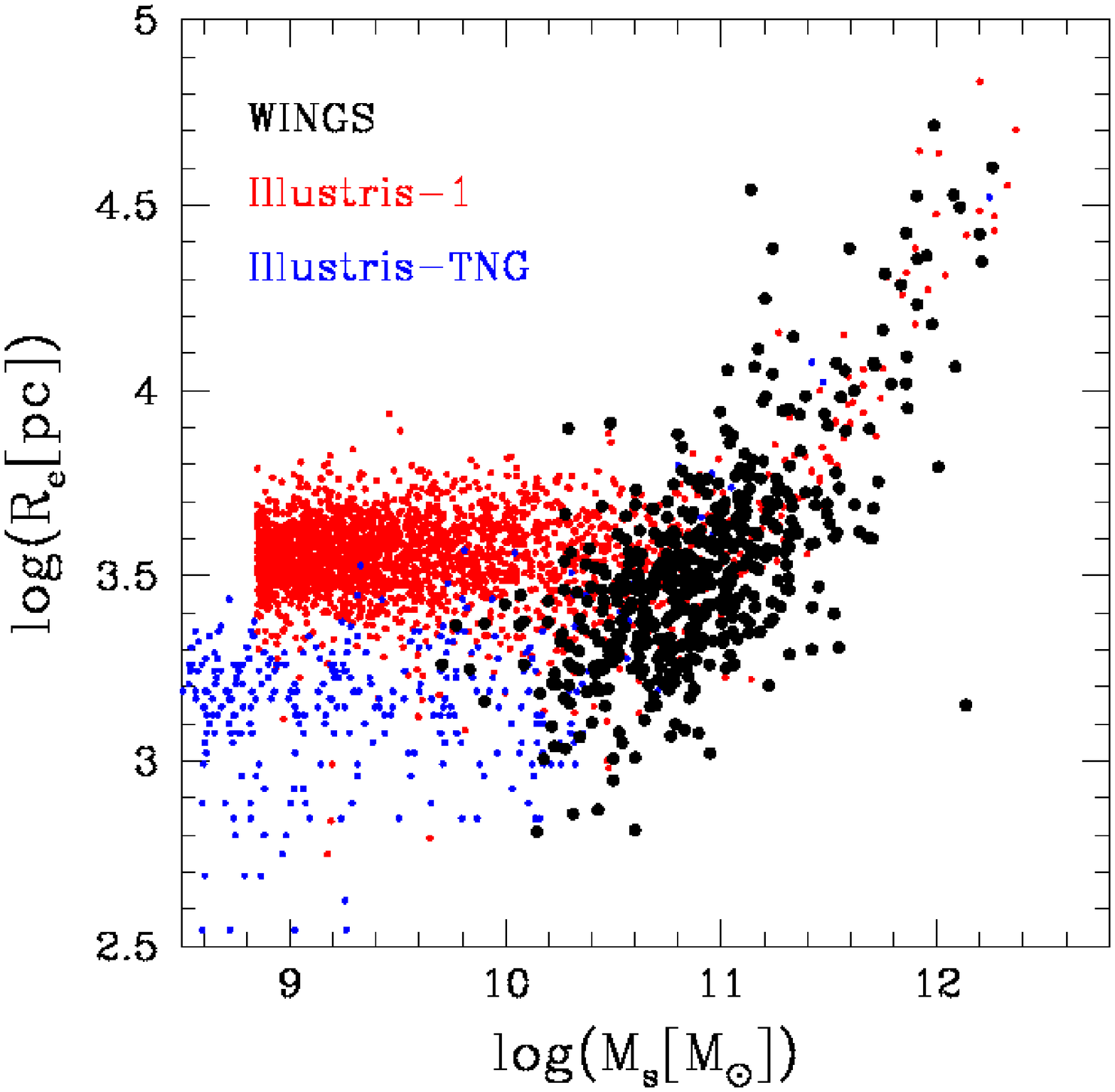}
      \includegraphics[scale=0.35]{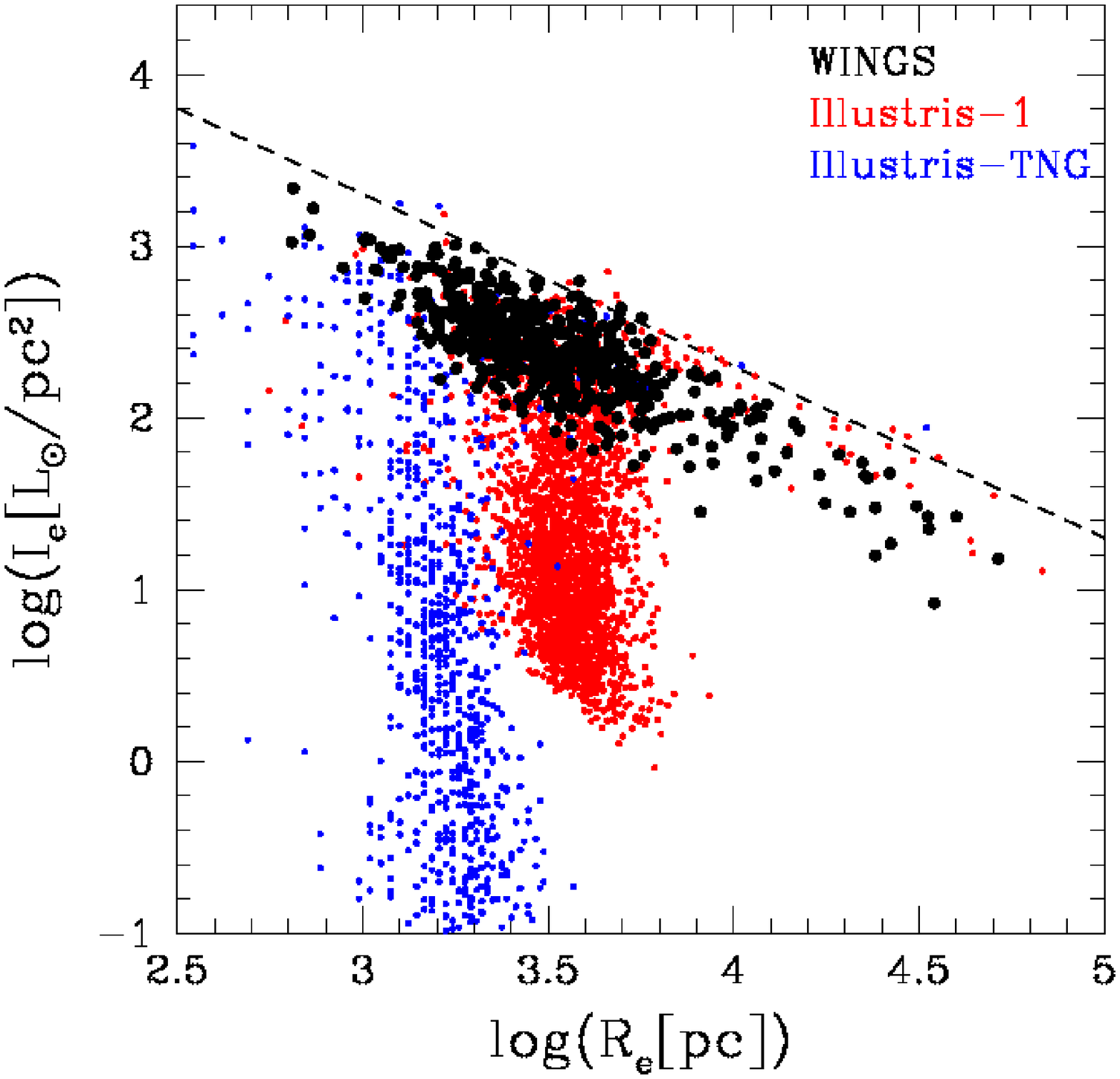}
         \includegraphics[scale=0.35]{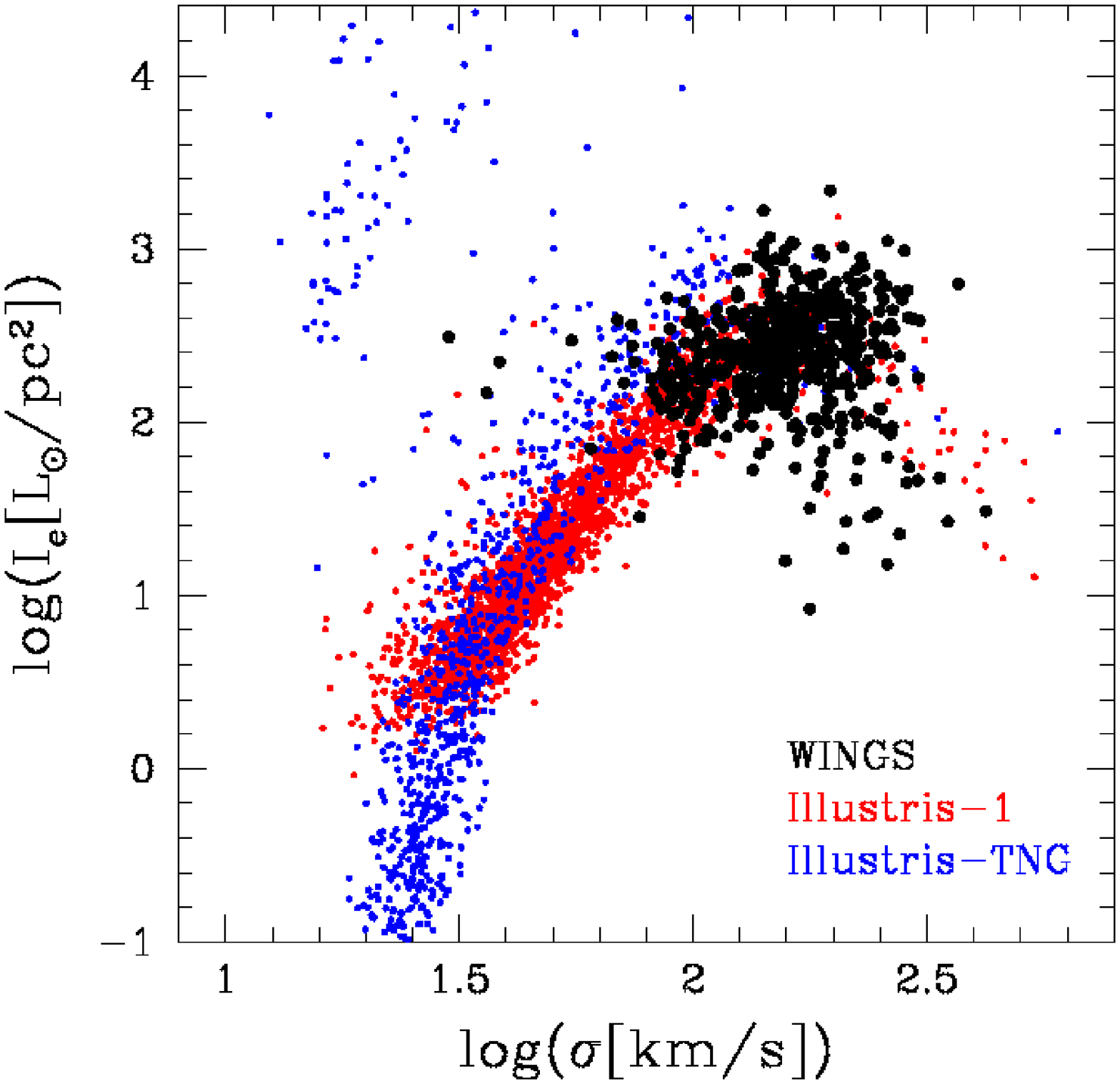}
            \includegraphics[scale=0.35]{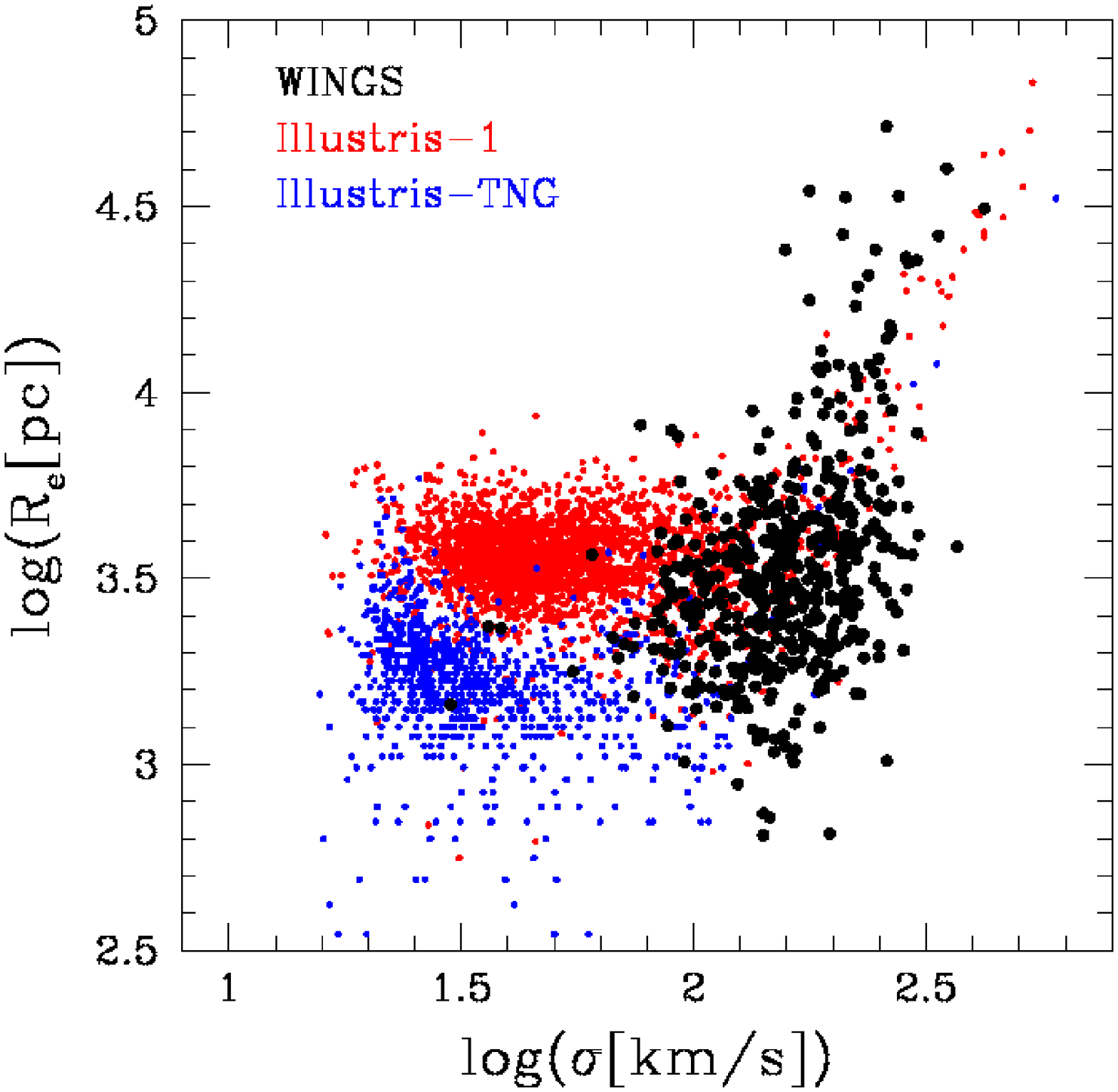}
   \caption{The \MRa\ plane and the three different projections of the FP. The black dots are the WINGS observational data. The red and blue dots are the data extracted from Illustris-1 and Illustris TNG-100 respectively. The galaxies with $\log(R_e)<2.5$  are not plotted here for better showing the bulk of the galaxy distribution.}
              \label{fig:3}
    \end{figure*}

{
Figure \ref{fig:3} shows the four main important SSRs for the WINGS and Illustris data. The left upper panel plots the stellar mass $M^*$ versus the effective radius \re\footnote{The symbols $M_s$ and $M^*$ both used in this work always refer to the total stellar mass in solar units.}. The WINGS data (black dots), the Illustris-1 data (red dots) and the Illustris-TNG data (blue dots) at $z=0$ are well visible.
We note that the \MRa\ relation is clearly non linear. The galaxies of small masses are distributed nearly horizontally, while the brightest galaxies follow a tail with a slope close to 1. The real and simulated data nicely superimpose each other over the same range of mass, even if the effective radii of Illustris-1 for the less luminous galaxies are systematically greater than the observational ones. In contrast, Illustris-TNG gives much smaller radii for the low mass galaxies. This is a well known fact already discussed in our papers of this series \citep[see \eg,][]{Donofrioetal2020, Donofrio_Chiosi_2022,Donofrio_Chiosi_2023}.
Both observations and simulations suggest the presence of the tail for the brightest ETGs, in which radii and masses are almost identical in observations and simulations. The different number of objects in the tail is due to different volumes sampled and to the way in which
the sample have been created: the WINGS and Illustris-1 datasets include only objects from clusters of galaxies, where large ETGs are frequent, while Illustris-TNG takes galaxies from the general field. In addition, the total volume of the surveys is different for WINGS, Illustris-1 and Illustris-TNG. 

The right upper panel of Fig. \ref{fig:3} shows the \IeRe\ plane obtained with the same data
(here the sample of WINGS galaxies is much smaller than in Fig. \ref{fig:2} because only the subsample is involved). Also in this case the most important fact to note is that the simulations correctly reproduce the presence of the tail for the brightest ETGs that is clearly separated from the cloudy distribution of the less luminous galaxies.
This tail, already seen in the original paper of \citet{Kormendy1977}, has a slope close to $-1$ (that predicted by the VT) and has been attributed to the peculiar evolution of the brightest galaxies that grow in mass by minor mergers \citep[among others, see \eg,][]{Capacciolietal1992}.  Our conclusion is therefore that both simulations catch the presence of some peculiar features of the \IeRe\ plane: the cloudy distribution of the faint galaxies, the tail formed by the brightest ETGs and the ZoE, \ie\ the region totally empty of galaxies above the dashed black line in Fig. \ref{fig:2}.

The lower panels of Fig. \ref{fig:3} show the other projections of the FP: the \IeSig\ and \Rsigma\ planes. Again we observe that both simulations are quite well superposed to the observational data. In particular the simulations are able to reproduce the curvature observed in the two distributions.

The good agreement between observations and simulations at $z=0$ is a good starting point. It tell us that simulations are able to reproduce the main features of the SSRs at $z=0$.
However, since our aim is to use simulations to infer the possible behavior of the SSRs at higher cosmic epochs, we need at least one further proof that simulations are able to catch the structural parameters of galaxies at much higher redshifts. To prove this we have used the data of \cite{DiSeregoetal2005}, who have analyzed the FP at $z\sim1$. In our Fig. \ref{fig:4} we can appreciate that the FP at this redshift epoch coming from the Illustris data (red and blue circles as before) is well in agreement with the observed one (black filled circles). The tilt of the plane is in practice identical with a value for the $a$ coefficient lower than 1 (for the Coma cluster the tilt provides $a\sim1.2$). The tilt is different with respect to that measured for the local clusters and this is an indication that there is an evolution of the structural parameters that seems to be reproduced by the simulations. Even in this case we note that the radii of the galaxies in the simulations are a bit systematically larger than those measured for the real galaxies, but this does not change the FP tilt of the simulated galaxies. Probably, since the total luminosity of the galaxies is quite well reproduced by the simulations, the combination of \Ie\ and \re\ is correct and the different effective radius simply change \Ie\ in such a way that the galaxy shift along the FP and not orthogonally to it.

%%%%%%%%%%%%Figure 4
   \begin{figure}
   \centering
   \includegraphics[scale=0.35]{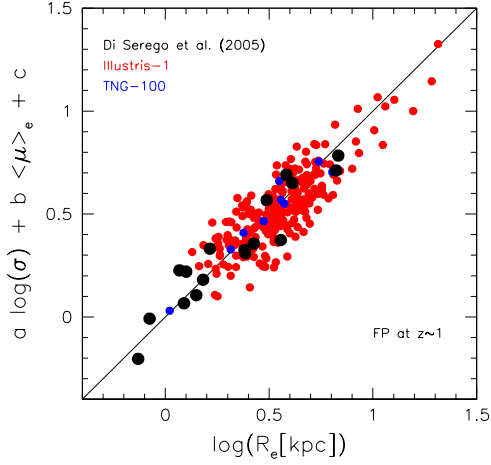}
   \caption{The FP at redshift $z=1$. The black filled circles are the data of \cite{DiSeregoetal2005}. The red and blue circles are for Illutris-1 and Illustris-TNG respectively. The FP has been determined considering only the galaxies with masses in the same range of the observational sample (those with $M^*>10^{10} M_\odot$).}
              \label{fig:4}
    \end{figure}

In concluding this section we observe that the artificial galaxies in the simulations are in quite good agreement with the real galaxies for what concern the main structural parameters even at much larger redshifts. The differences with real galaxies mainly concern the stellar content, the colors and the star formation rates, but these differences do not seem to plague the general behavior of the SSRs. For this reason we believe that it is possible to extract some information on the evolution of galaxies, by looking at the distributions of galaxies in the SSRs. When high redshift data will be available in good number, we could better compare observations and simulations and extract useful information on the evolution of galaxies.
}

\section{The basic equations of our framework}
\label{sec:3}

Before starting the discussion of the main  SSRs predicted for the most far cosmic epochs, it is important to summarize here the main conclusions drawn by \cite{Donofrio_Chiosi_2022,Donofrio_Chiosi_2023} using the combination of the VT and the \Lsigb\ law\footnote{From here on we drop the time notation for simplicity.}. This combination is the key novelty of their approach and a necessary premise for understanding what follows. The two equations representing the VT and the \Lsigb\ law are:

%%%%%%%%%%%%%%%%equation 1
\begin{eqnarray}
 \sigma^2 &= & \frac{G}{k_v} \frac{M_s}{R_e}  \nonumber \\
 \sigma^\beta &= & \frac{L}{L'_0} = \frac{2\pi I_e R^2_e}{L'_0}. 
\label{eqsig}
\end{eqnarray}

In these equations $\beta$ and $L'_0$ are free time-dependent parameters that depend on the peculiar history of each object. 
From these equations one can derive all the mutual relationships existing among the parameters $M_s$, $R_e$, $ L$, $I_e$, $\sigma$ characterizing a galaxy. We find:

\begin{equation}
I_e  = \Pi R_e^{\gamma}
\label{eqIeRe}
\end{equation}
\noindent
for the \IeRe\ plane, where 

$$\gamma=\frac{(2/\beta)-(1/2)}{(1/2)-(1/\beta)}$$

\noindent
and $\Pi$ is a factor that depends on $k_v$, $M/L$, $\beta$, and $L'_0$. It is given by

$$
\Pi  = \left [ \left (\frac{2\pi}{L'_0}\right )^{1/\beta} \left (\frac{L}{M_s} \right )^{(1/2)} \left (\frac{k_v}{2\pi G} \right )^{(1/2)} \right ]^{\frac{1}{1/2-1/\beta}}.
\label{eq4}
$$
Then we have:
%%%%%%%%%%%%%%%%equation 13
\begin{equation}
    I_e = \left [ \frac{G}{k_v}\frac{L'_0}{2\pi}M_s \Pi^{3/\gamma} \right ]^{\frac{\beta-2}{1+3/\gamma}} \sigma^{\frac{\beta-2}{1+3/\gamma}}
    \label{eqIeSig}
\end{equation}
for the \IeSig\ relation and
%%%%%%%%%%%%%%%%equation 14
\begin{equation}
    R_e = \left [ \frac{G}{k_v}\frac{L'_0}{2\pi}\frac{M_s}{\Pi} \right ] \sigma^{\frac{\beta-2}{3+\gamma}}
    \label{eqReSig}
\end{equation}
for the \Rsigma\ relation.
In addition we have:
%%%%%%%%%%%%%%%%equation 15
\begin{equation}
    R_e = \left [ (\frac{G}{k_v})^{\beta/2} \frac{L'_0}{2\pi} \frac{1}{\Pi} \right ]^{\frac{2(\beta-2)}{\beta^2-6\beta+12}} M_{s}^{\frac{\beta^2-2\beta}{\beta^2-6\beta+12}}
    \label{eqReM}
\end{equation}
for the \MRa\ relation.

It is important to note here that in all these equations the slopes of the log relations depend only on $\beta$. This means that when a galaxy changes its luminosity $L$ and its velocity dispersion $\sigma$, \ie\ when $\beta$ has a well defined value (either positive or negative), the effects of the motion in the \Lsig\ plane are propagated in all the FP projections. In these planes the galaxies cannot move in whatever directions, but are forced to move only along the directions (slopes) predicted by the $\beta$ parameter in the above equations. In this sense the $\beta$ parameter is the link we are looking for between the FJ (and the FP) and the observed distributions in the FP projections.

In addition, the combination of eqs. (\ref{eqsig}) gives us another important equation. It is now possible to write a FP-like equation valid for each galaxy depending on the $\beta$ and $L'_0$ parameters:

%%%%%%%%%%%%%%%%equation 2
\begin{equation}
    \log R_e = a \log\sigma + b <\mu>_e + c
    \label{eqfege}
\end{equation}

\noindent
where the coefficients:

\begin{eqnarray}
a & = & (2+\beta)/3 \\ \nonumber
b & = & 0.26 \\ \nonumber
c & = & -10.0432+0.333*(-\log (G/k_v) - \log (M/L) \\ \nonumber
   &   & -2*\log (2\pi)-\log (L'_0)) \nonumber
\end{eqnarray}

\noindent
are written in terms of $\beta$ and $L'_0$. We note that this is the equation of a plane whose slope depends on $\beta$ and the zero-point on $L'_0$. The similarity with the FP equation is clear. The novelty is that the FP is an equation derived from the fit of a distribution of real objects, while here each galaxy independently follows an equation formally identical to the classical FP but of profoundly different physical meaning. In this case, since $\beta$ and $L'_0$ are time dependent, the equation represents the instantaneous plane on which a generic galaxy is located in the \FPR\ space and consequently in all its projections.

Finally, the combination of the above equations allows us  to determine the values of $\beta$ and $L'_0$, the two critical evolutionary parameters. This is possible by writing the following equations:

%%%%%%%%%%%%%%%%equation 8
\begin{eqnarray}
\beta [\log(I_e)+\log(G/k_v)+\log(M_s/L)+\log(2\pi)+\log(R_e)] + \\ \nonumber
    + 2\log(L'_0) - 2\log(2\pi) - 4\log(R_e) = 0 \\ 
 \beta\log(\sigma) + \log(L'_0) + 2\log(\sigma) + \log(k_v/G) - \log(M_s) + \\ \nonumber
 - \log(2\pi) - \log(I_e) - \log(R_e) = 0. 
\label{eqbet}
\end{eqnarray}
\noindent
{  Posing now: }

%%%%%%%%%%%%%%%%equation 9
\begin{eqnarray}
A  & = & \log(I_e)+\log(G/k_v)+\log(M_s/L)+\log(2\pi)+ \\ \nonumber
   &   & \log(R_e)  \\ \nonumber
B  & = & - 2\log(2\pi) - 4\log(R_e)  \\ \nonumber
A' & = &  \log(\sigma)  \\ \nonumber
B' & = & 2\log(\sigma) - \log(G/k_v) - \log(M_s) - \log(2\pi) -  \\ \nonumber 
   &   & \log(I_e) - \log(R_e)  \\ \nonumber
\end{eqnarray}
{  we obtain} the following system:

%%%%%%%%%%%%%%%%equation 10
\begin{eqnarray}
A\beta + 2\log(L'_0) + B = 0 \\
A'\beta + \log(L'_0) + B'= 0
\label{eqsyst}
\end{eqnarray}
\noindent 
with solutions:

%%%%%%%%%%%%%%%%equation 11
\begin{eqnarray}
 \beta & = & \frac{-2\log(L'_0) - B}{A} \\ 
 \log(L'_0) & = &\frac{A'B/A - B'}{1-2A'/A}.
 \label{eqbeta}
\end{eqnarray}

The key result is that the  parameters $L$, $M_s$, \re, \Ie\ and $\sigma$ of a galaxy fully determine the evolution that is encoded in the parameters $\beta$ and $L'_0$.

Considering the fact that each structural parameter is known with a maximum error of $\sim20$\%, the single values of beta cannot be trusted too much. On average instead we will show that the galaxies  move in the SSRs only in the directions defined by beta.

Given this premise,  we proceed now to show the basic SSRs at much higher redshifts. 

 %%%%%%%%%%%%Figure 5
   \begin{figure*}
   \centering
   \includegraphics[angle=-90, scale=0.65]{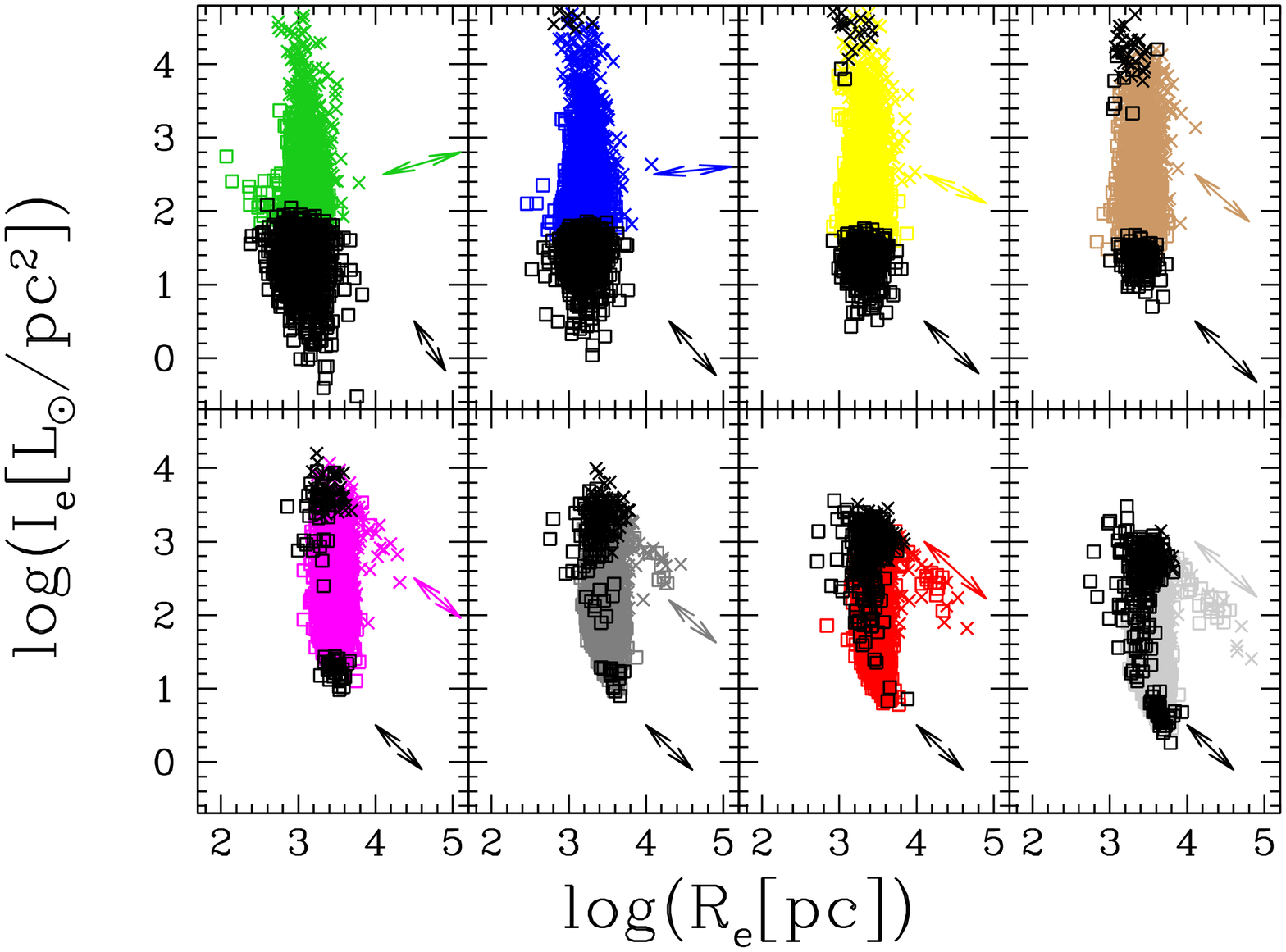} 
   \caption{The \IeRe\ plane at different redshift for Illutris-1. From top left to bottom right we can see the distribution of galaxies at $z=4$, $z=3$, $z=2.2$, $z=1.6$, $z=1.0$, $z=0.6$, $z=0.2$ and $z=0$. The colored dots mark the galaxies with $\beta>0$ at each epoch, whereas the black symbols those with $\beta\leq0$. The crosses mark the galaxies with SFR greater than the mean <SFR> at each epoch, while the open squares those with SFR lower than <SFR>. The arrows indicate the average direction of motion predicted on the basis of the values of $\beta$ from eq. (\ref{eqIeRe}) (see also Table \ref{beta_values}). The colored arrows mark the average for the galaxies with $\beta>0$ and the black arrows those with $\beta\leq0$. Their length is not related or proportional to any other quantity: it has been chosen for graphical reasons.}
              \label{fig:5}
    \end{figure*}

%%%%%%%%%%%%Figure 6
   \begin{figure*}
   \centering
   \includegraphics[angle=-90, scale=0.65]{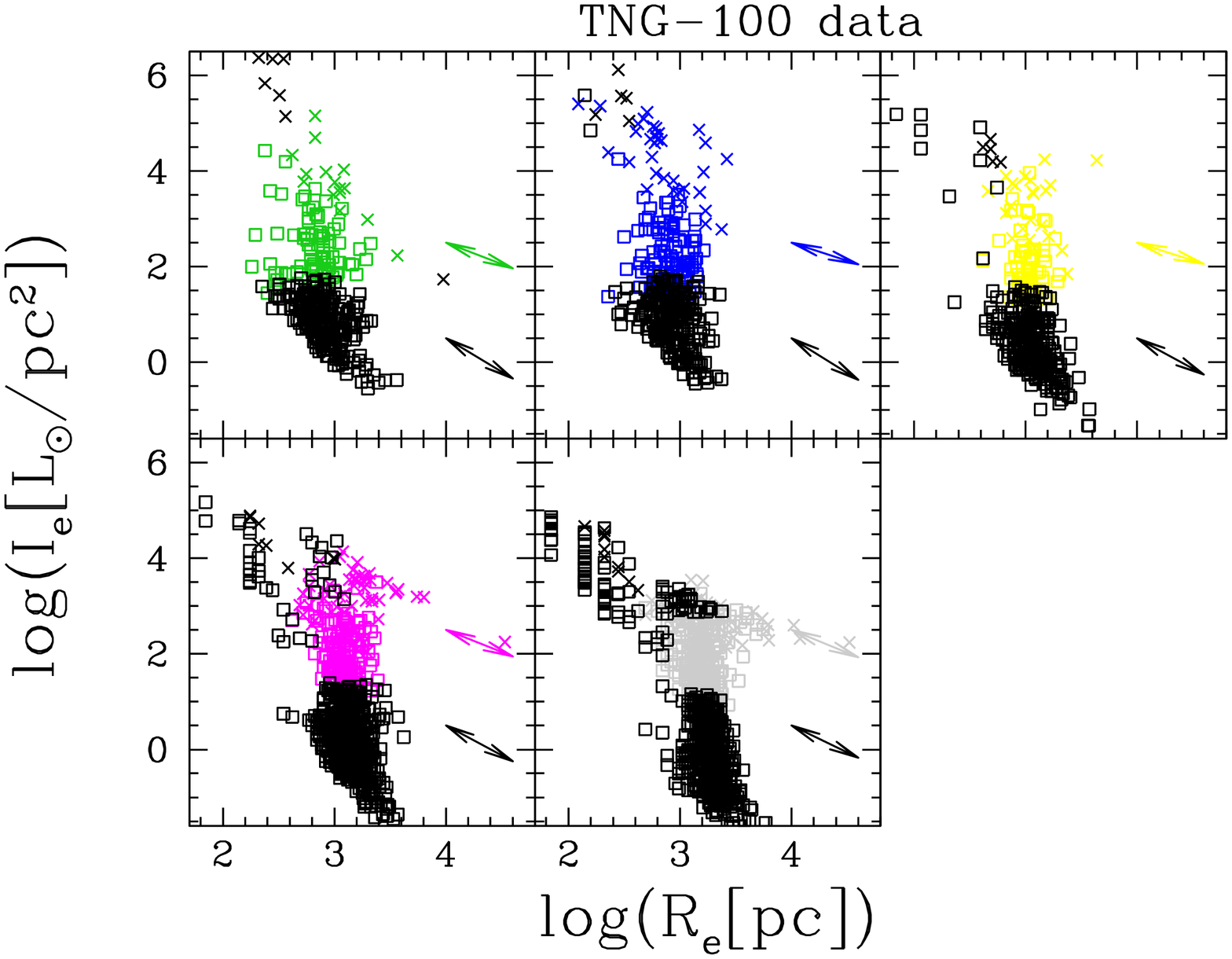} 
   \caption{The \IeRe\ plane at different redshift for the TNG-100 data. From top left to bottom right we can see the distribution of galaxies at $z=4$, $z=3$, $z=2.2$, $z=1.0$, and $z=0$. The colored dots mark the galaxies with $\beta>0$ at each epoch, whereas the black symbols those with $\beta\leq0$. The crosses mark the galaxies with SFR greater than the mean <SFR> at each epoch, while the open squares those with SFR lower than <SFR>. The arrows indicate the average direction of motion predicted on the basis of the values of $\beta$ from eq. (\ref{eqIeRe}) (see also Table \ref{beta_values}). The colored arrows mark the average for the galaxies with $\beta>0$ and the black arrows those with $\beta\leq0$. Their length is not related or proportional to any other quantity: it has been chosen for graphical reasons.}
              \label{fig:6}
    \end{figure*}

\section{The  SSRs at high redshift}
\label{sec:4}
    
To explore the behavior of the  SSRs at  high redshifts we can only relay on simulations because we do not have enough observational data for galaxies at high redshift. 
Fortunately, despite the small systematic overestimate of the effective radii, the Illustris-1 and Illustris-TNG data are sufficiently good to be trusted even at high redshifts. Furthermore, as shown by \citet{Donofrio_Chiosi_2023}, both Illustris-1 and Illustris-TNG produce a very similar distribution for the $\beta$ parameter. 
Thanks to this, the simulated data can provide a reliable insight  on the evolution of the SSRs with time. In the following we will show the results for both the Illustris-1 and Illustris-TNG samples currently available to us. 

{
Figures \ref{fig:5} and \ref{fig:6} present the \IeRe\ plane from $z=4$ (upper left) to $z=0$ (bottom right) for the Illustris-1 and Illustris-TNG respectively. For Illustris-1 the whole sequence of redshifts is $z=4$ (left upper panel), $z=3$, $z=2.2$, $z=1.6$, $z=1.0$, $z=0.6$, $0.2$, $z=0$ (right bottom panel) as indicated.   In all the panels, the colored  dots indicate galaxies with $\beta > 0$ and the black points those with $\beta < 0$. Crosses and open squares indicate galaxies with SFR greater than the average <SFR> or lower than the average respectively at each redshift epoch.
The same color code is also used for the arrows indicating the mean slope of $\beta$, calculated from eq. (\ref{eqbeta}) for each object of the simulation). Such mean value provides approximately the direction of motion in this plane for most of the galaxies as expected from eq. (\ref{eqIeRe}).

The sequence of panels indicates that the tails well visible at $z=0$ for the brightest galaxies start to appear at $z\sim 1-1.5$. This epoch probably corresponds to the time  in which minor  mergers either with or without star formation   on already formed massive objects became the typical event  thus increasing both the mass and radius of galaxies  \citep{Naabetal2009}. It is interesting to note that the directions of the arrows, whose slope depends on eq. (\ref{eqIeRe}) (see Table \ref{beta_values}), {flip progressively with $z$}, in particular for the positive $\beta$'s, assuming the value close to $-1$ (as predicted by the VT) approximately at $z\sim 0.6$, and remaining constant thereafter.
Such slope gives the only possible direction of motion of galaxies in the \IeRe\ plane at each epoch when the evolution proceeds and $\beta$ changes. The brightest galaxies, that have likely reached a full virial equilibrium far in the past, are no longer affected by strong episodes of star formation, and start to move along this direction at $z\sim 1-1.5$, forming the tail we observe today. Notably, even the galaxies with $\beta\leq 0$ progressively reach the same slope. This happens because several objects have large negative $\beta$ values. As we will see later, both positive and negative values of $\beta$ are possible and, as demonstrated by \cite{DonofrioChiosi2021}, this is a necessary condition for reproducing the \Lsig\ distribution starting from the \IeRe\ distribution (and viceversa). 

A further thing to note is that the galaxies with strong SFR (greater than <SFR>) have in general a positive $\beta$. When some object with negative $\beta$ appears on top of the distribution, it has a SFR > <SFR>. Only later on, when the present epoch is approached, we start to see in the upper part of the cloud, galaxies with negative $\beta$ and SFR < <SFR> (the black open squares). These objects might be relatively small compact galaxies where the SF is over (a possible candidate for this class of galaxies could be M32). Notably the galaxies with the higher surface brightness have positive $\beta$ at high redshift and only later this region of the plot is populated by objects with negative $\beta$ and low SFR.}

%%%%%%%%%%%%Figure 7
   \begin{figure*}
   \centering
   \includegraphics[angle=-90, scale=0.65]{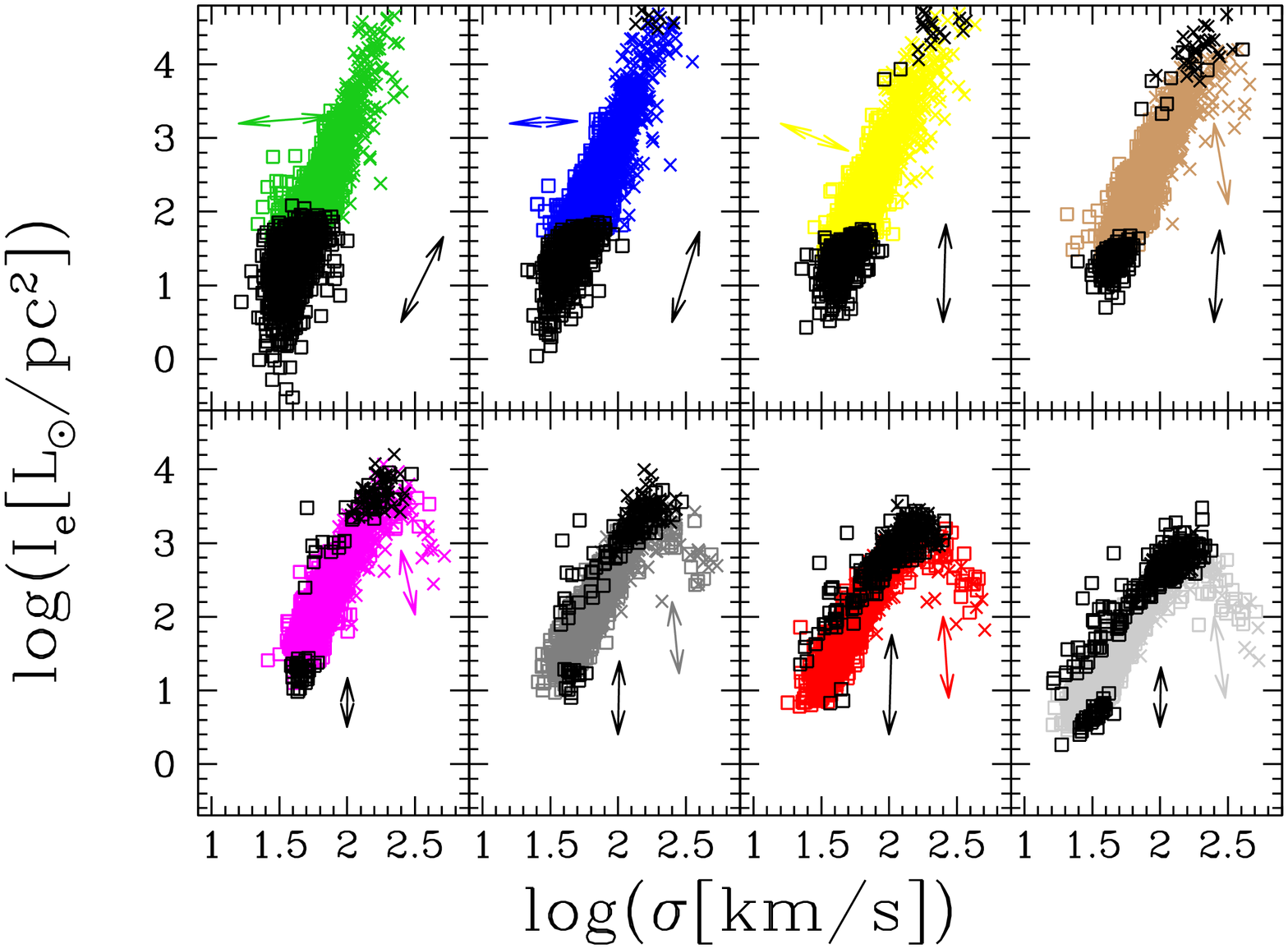}
   \caption{The \IeSig\ plane for the Illustris-1 data. Symbols and colors as in Fig. \ref{fig:5}.}
              \label{fig:7}
    \end{figure*}

%%%%%%%%%%%%Figure 8
   \begin{figure*}
   \centering
   \includegraphics[angle=-90, scale=0.65]{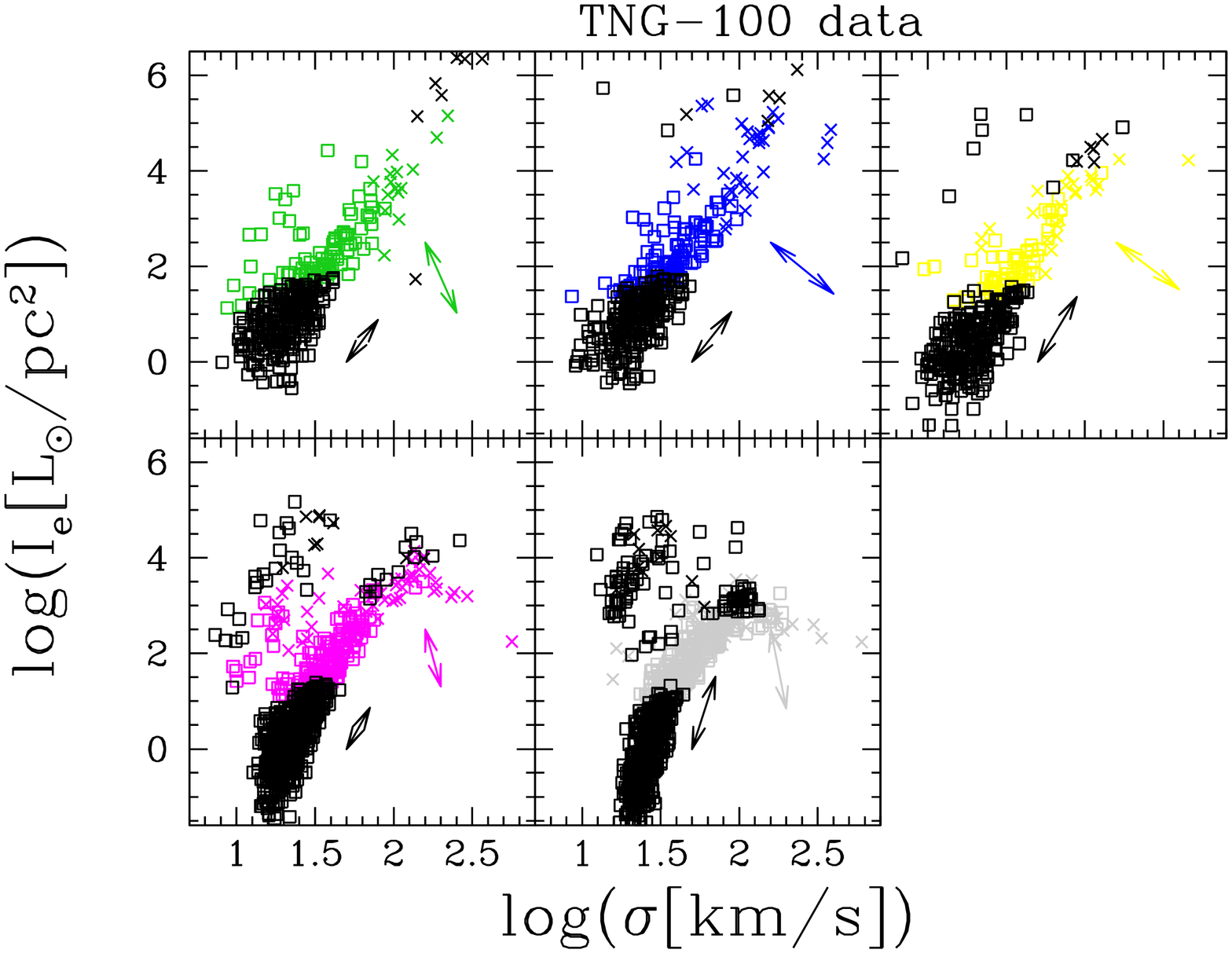}
   \caption{The \IeSig\ plane for the Illustris-TNG data. Symbols and colors as in Fig. \ref{fig:6}.}
              \label{fig:8}
    \end{figure*}
    
{A very similar behavior is observed in Fig. \ref{fig:6} when we use the Illustris-TNG data. As in Fig. \ref{fig:5} the colored dots indicate the galaxies with positive $\beta$ and the arrows the mean value of $\beta$. With respect to the previous figure we note that the colored arrows do not change very much their directions. We attribute this behavior to the different dimension of the two samples.
For the Illutris-TNG data the arrows have always slope close to $-1$, a fact that indicate quite big values for $\beta$. In any case, the trend of populating the upper region of high surface brightness with object with negative $\beta$ and low <SFR> is confirmed.}

One might ask now what produces the tail for the brightest objects in the \IeRe\ plane, being the arrows directed at all redshift epochs approximately toward the same directions. This can be better understood looking at the other FP projections and 
taking into account that the physical mechanisms at work in small and large galaxies can be different.

Figure \ref{fig:7} is even more impressing in showing how the changes of $\beta$ across time determine the motions in the FP projections  (the symbols and color code have the same meaning of those in Fig.\ref{fig:5}).
In the \IeSig\ plane  the curvature formed by the brightest galaxies is much more pronounced. We note again that as $\beta$ increases, the colored arrows point to the direction of the tail that we see at $z\sim 0$. In the figure we also note that the galaxies with negative $\beta$ are preferentially at the bottom of the cloud distribution and their number decreases up to $z\sim0.6-1.0$. Then they increase again in number and tend to crowd the top region of the distribution. {As before the upper region of the distribution with high \Ie\ is populated by objects with high SFR at the most remote epochs and only approaching $z=0$, galaxies with negative $\beta$ and low SFR appears on top of the distribution.}

%%%%%%%%%%%%Figure 9
   \begin{figure*}
   \centering
   \includegraphics[angle=-90, scale=0.65]{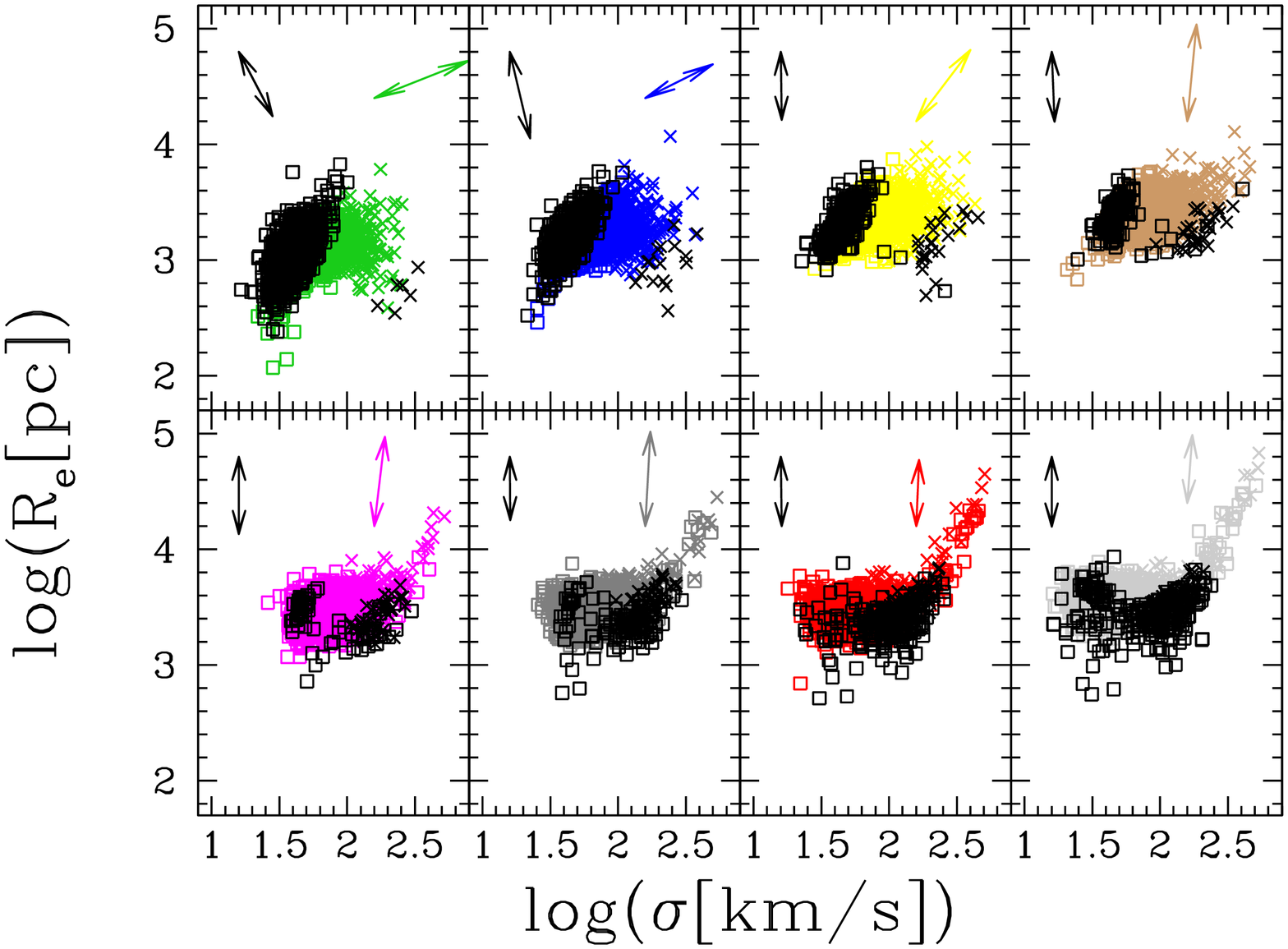}
   \caption{The \Rsigma\ plane for Illutris-1. Symbols and colors as in Fig. \ref{fig:6}.}
              \label{fig:9}
    \end{figure*}

%%%%%%%%%%%%Figure 10
   \begin{figure*}
   \centering
   \includegraphics[angle=-90, scale=0.65]{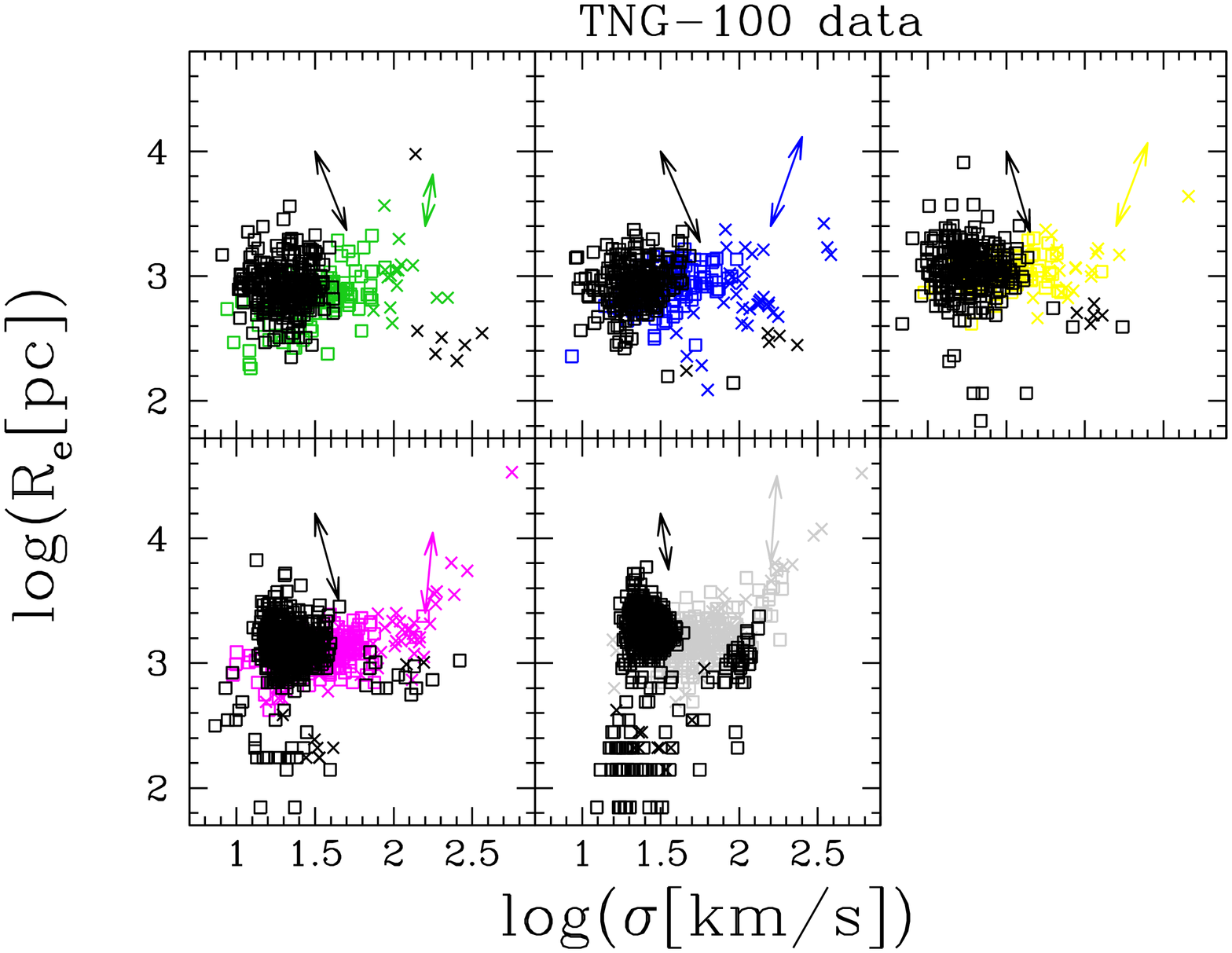}
   \caption{The \Rsigma\ plane for Illustris-TNG. Symbols and colors as in Fig. \ref{fig:7}.}
              \label{fig:10}
    \end{figure*}
    
Again the TNG data (Fig. \ref{fig:8}) give a similar picture of the \IeSig\ plane. Now the change of direction due to positive and negative values of $\beta$ is much evident and we understand that the tail originates when $\beta$ increases and the galaxies progressively become much virialized.

Figures \ref{fig:9} and \ref{fig:10} display the \Rsigma\ plane with the Illutris-1 and Illutris-TNG data, respectively. Again the slopes of the arrows predicted by eqs. (\ref{eqIeRe}), (\ref{eqIeSig}), and (\ref{eqReSig}) are in good agreement with those inferred from the observed distribution of real galaxies and explain the tail formed by the bright galaxies.

\begin{figure*}
   \centering
   \includegraphics[angle=-90, scale=0.65]{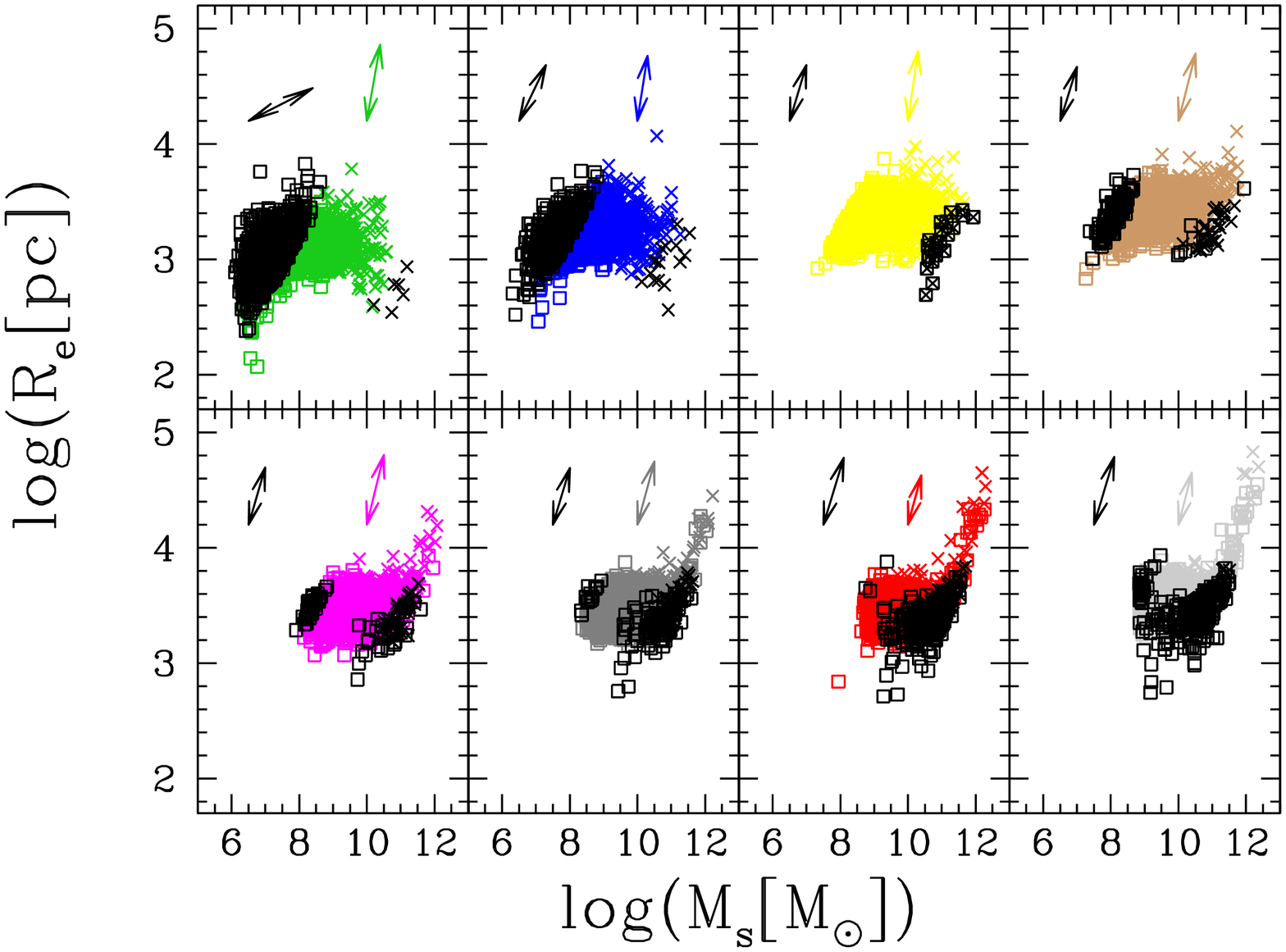}
   \caption{The \MRa\ plane for Illustris-1. Symbols and colors as in Fig. \ref{fig:5}.}
              \label{fig:11}
    \end{figure*}

\begin{figure*}
   \centering
   \includegraphics[angle=-90, scale=0.65]{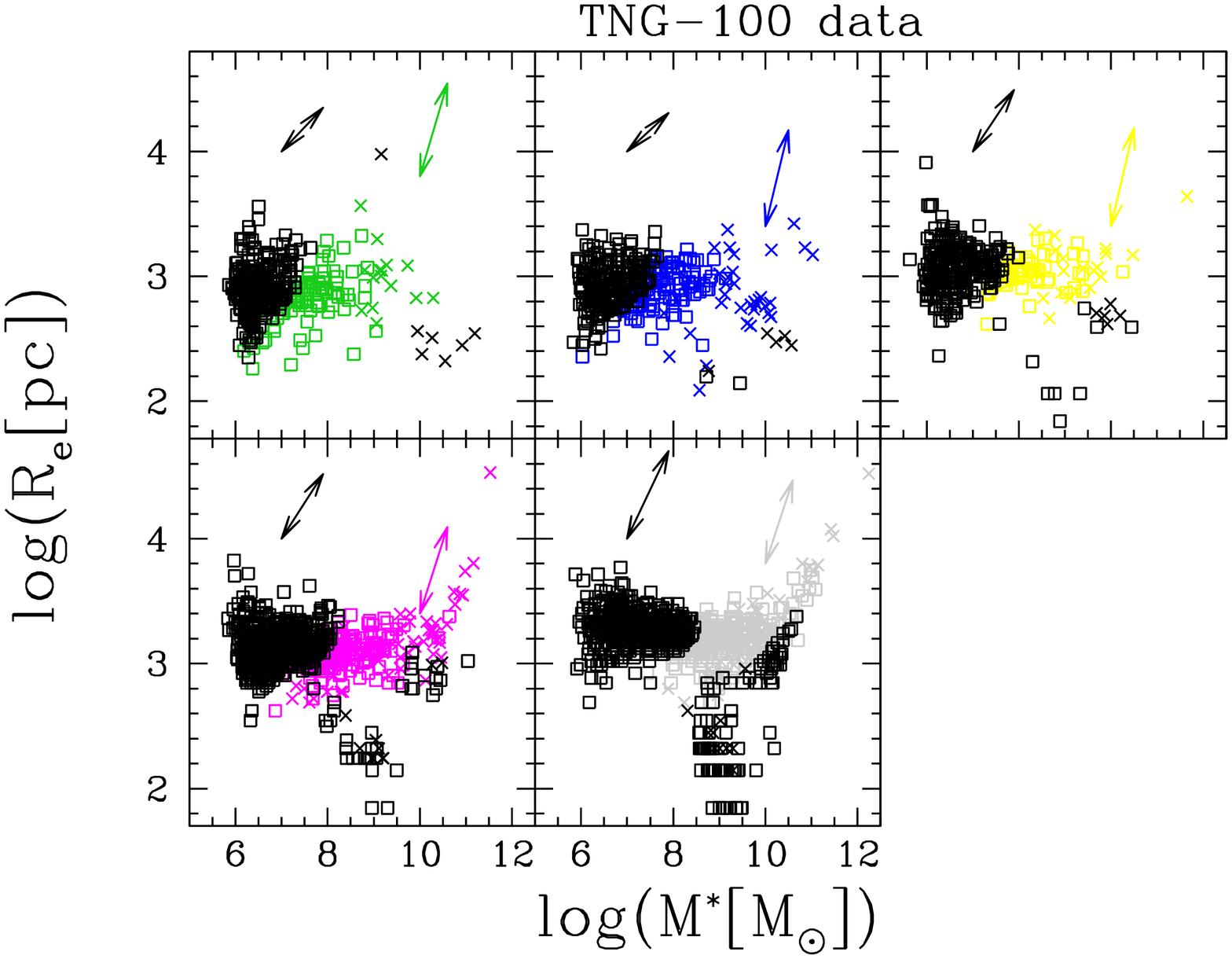}
   \caption{The \MRa\ plane for Illustris-TNG. Symbols and colors as in Fig. \ref{fig:6}.}
              \label{fig:12}
    \end{figure*}

The same can be said for the \MRa\ plane (Figs. \ref{fig:11} and \ref{fig:12}). In both  planes the tail formed by the brightest galaxies stands out clearly. 
The slope of the tail in the \MRa\ plane is exactly close to 1,  as predicted by the VT. The last bottom right panel (that at $z=0$) shows in particular that objects both with negative and positive large values of $\beta$ begin to climb the tail as soon as their mass exceeds about $10^{11} \, M_\odot$.

As in the \IeRe\ plane, the TNG data exhibit quite similar mean values for the slopes predicted at the different redshifts. The slope is close to 1 and this means that the majority of the galaxies are quite well virialized at $z=0$. {A further notable thing is that the TNG100 data indicate the presence of quite massive objects with small radii, not visible in Illustris-1. These objects might be the class of compact massive galaxies with high \Ie\ also visible in Fig. \ref{fig:6} and \ref{fig:8}. Notably we can see that all these objects have negative $\beta$ and very low SFR. In other words they are isolated compact massive galaxies where the SF stopped long time ago.}

The picture we have illustrated here using the Illustris-1 and Illutris-TNG data clearly reveals a progressive trend of the galaxies toward the full virial equilibrium as indicated by the slopes of the arrows when $|\beta| \rightarrow \infty$. Such condition is reached by the most massive galaxies approximately at $z\sim1.5-1.0$.
In general the galaxies with the largest radii have $\beta>0$ both in simulations and observations. This behavior is compatible with the predictions of minor mergers in which galaxies might increase their radius without changing significantly their mass and luminosity \citep{Naabetal2009,Geneletal2018}.

%%%%%%%%%%%%Figure 11

%%%%%%%%%%%%Figure 12

\begin{table*}
\begin{center}
\caption{The slopes of the \IeRe, $R_e-\sigma$, \IeSig\ and \MRa\ planes for different values of $\beta$.}
                \label{beta_values}
                \begin{tabular}{|r| r| r| r| r| r| r| r|}
\hline
         $\beta$     & \IeRe\ & $R_e-\sigma$ $^a$ & \IeSig\ $^b$ & $R_e$-$M_s$ $^c$ &  $R_e-\sigma$ $^d$ & \IeSig\ $^e$ & $R_e$-$M_s$ $^f$  \\
\hline
       100.0 &      -0.98   &    102.0  &  98.0 &  0.96 & 48.50 & -47.51 & 1.04 \\
        50.0 &      -0.96   &     52.0  &  48.0 &  0.92 & 23.51 & -22.53 & 1.08 \\
        10.0 &      -0.75   &     12.0  &   8.0 &  0.71 &  3.55 &  -2.66 & 1.54 \\
         5.0 &      -0.33   &      7.0  &   3.0 &  0.55 &  1.12 &  -0.37 & 2.14 \\
         3.0 &       1.00   &      5.0  &   1.0 &  0.43 &  0.25 &   0.25 & 1.00 \\
         2.0 &       0.00   &      4.0  &   0.0 &  0.33 &  0.0  &   0.0  & 0.00 \\
         1.0 &      -3.00   &      3.0  &  -1.0 &  0.20 &  0.0  &   0.0  &-0.14 \\
         0.5 &      -2.33   &      2.5  &  -1.5 &  0.11 & -2.25 &   5.25 &-0.08 \\
         0.0 &      -2.00   &      2.0  &  -2.0 &  0.00 & -2.00 &   4.00 & 0.28 \\
        -0.5 &      -1.80   &      1.5  &  -2.5 & -0.14 & -2.08 &   3.74 & 0.08 \\
        -1.0 &      -1.67   &      1.0  &  -3.0 & -0.33 & -2.25 &   3.75 & 0.16 \\
        -2.0 &      -1.50   &      0.0  &  -4.0 & -1.00 & -2.66 &   4.00 & 0.28 \\
        -3.0 &      -1.40   &     -1.0  &  -5.0 & -3.00 & -3.12 &   4.37 & 0.38 \\
        -5.0 &      -1.28   &     -3.0  &  -7.0 &  5.00 & -4.08 &   5.25 & 0.52 \\
       -10.0 &      -1.16   &     -8.0  & -12.0 &  1.67 & -6.54 &   7.63 & 0.69 \\
       -50.0 &      -1.03   &    -48.0  & -52.0 &  1.08 &-26.51 &  27.53 & 0.92 \\
      -100.0 &      -1.02   &    -98.0  & -102.0&  1.04 &-51.50 &  52.51 & 0.96 \\
\hline
\end{tabular}
\end{center}
Notes: a) Slope when $k_v$, $M_s$ and $I_e$ are constant; b) Slope when $k_v$, $M_s$ and $R_e$ are constant; c) Slope when $k_v$, and $I_e$ are constant; d) Slope when $k_v$ and $M_s$ are constant: e) Slope when $k_v$ and $M_s$ are constant; f) Slope when $k_v$ is constant.
\end{table*}

Finally, we have to remark that the FJ relation does not change very much with redshift. As the redshift decreases from $z=4$ to $z=0$ through six intermediate steps, we observe that some galaxies have negative $\beta$ and others have positive $\beta$. The fits of the observed distributions reveal that the slope of the FJ relation progressively decreases passing from nearly 4 (at $z=4$) to nearly 2 (at $z=0$). 

{
One may legitimately ask why the scatter of the FJ relation does non increase with time if there are objects that move nearly perpendicularly to the trend indicated by the observed distribution. The same trend is visible with the TNG data (not plotted here).

We believe that the scatter cannot increase as consequence of the merger activity because the maximum possible variation in luminosity that a galaxy might experience  does not exceed a factor of two (when a galaxy approximately doubles its mass merging with a similar object of the same mass and stellar content) that  in log units corresponds to a factor of $\sim 0.3$, a very small shift compared with the scale spanned by the data values. 

To support this statement in Appendix \ref{appendix_A} we present a toy model predicting  the effects   on the total luminosity of a galaxy with mass $M_1$, age $T_1$ and luminosity $L_1$  would undergo as a consequence of a merger with another object of mass $M_2$, age $T_2$ and luminosity $L_2$.  The event may or may not  be followed  by star formation engaging a certain amount of gas with mass $M_3$. 

Using reasonable values for the masses and luminosities of the three components (see eq.(\ref{eq_merger_burst} in Appendix \ref{appendix_A} ) we may expect that  
the total luminosity first increases and then decreases on a timescale that depends on the amount of matter engaged in the burst of activity. In any case the luminosity evolution is fast up to a few $10^8$ years after the burst (turnoff mass about 3$ M_\odot$) slows down up to $10^9$ years (turnoff mass about 2 $M_\odot$), and then becomes even slower afterwards. The estimated fading rate of the luminosity is about $|\Delta (logL/L_\odot)| \simeq 0.015$ per Gyr and per unit SSP mass, must be multiplied by 5.8 to get the real fading rate per Gyr \citep[see the SSP database of][]{Tantalo_2005}.  Consequently it is very unlikely to catch a galaxy exactly at the time of the maximum luminosity. Equation (\ref{eq_merger_burst}) allows us to quickly evaluate the effects of mergers with different combinations of masses and ages of the involved galaxies. However, the examples shown in Appendix \ref{appendix_A}  demonstrate that but for the case of a merger between to objects of comparable mass, in which the luminosity and mass of the resulting objects are double the original ones, mergers among objects of different mass, age and likely undergoing some star formation during the merger generate  objects that  in practice do not keep trace of the merger but simply keep the properties (mass and luminosity) of the most massive component.  More details are not of interest here. }

%%%%%%%%%%%%Figure 13
   \begin{figure}
   \centering
   \includegraphics[scale=0.40]{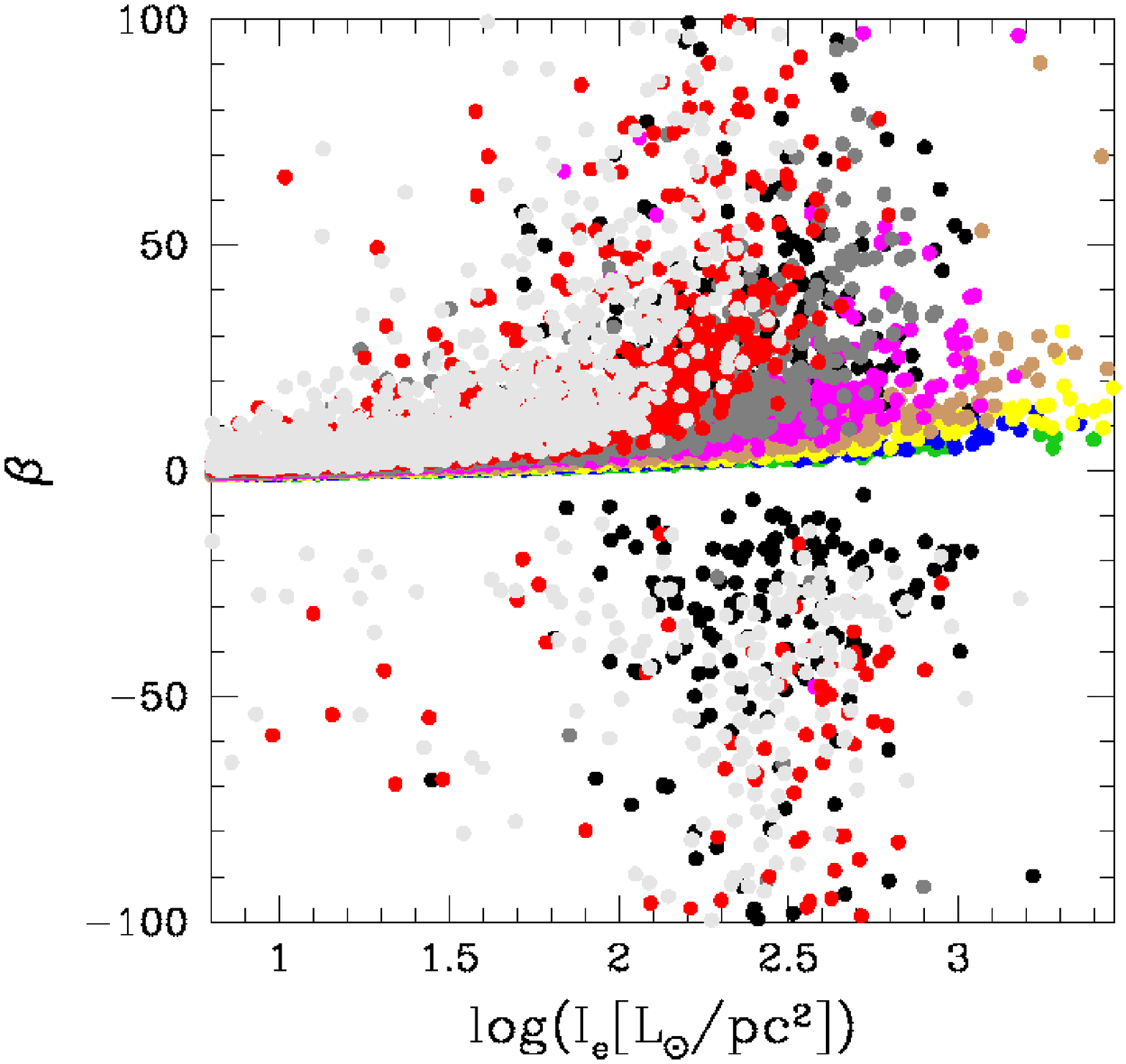}   
   \caption{The $\beta-\log(I_e)$ plane. The color dots have the same code of that in Fig. \ref{fig:5}, marking the galaxies at different redshift epochs. The black dots are the data of the WINGS database at $z=0$.}
              \label{fig:13}
    \end{figure}

The main conclusion of this section can be summarized as follows. The hypothesis that the VT and the \Lsigb\ work together to govern the evolution of mass $M_s$, luminosity $L$, radius $R_e$, surface brightness $I_e$ and velocity dispersion $\sigma$, leads to a coherent and self-consistent explanation of all the scale relations of galaxies together with a reasonable explanation for the tilt of the FP as demonstrated by  \cite{Donofrio_Chiosi_2022,Donofrio_Chiosi_2023}.

%%%%%%%%%%%%Figure 14
   \begin{figure}
   \centering
   \includegraphics[scale=0.40]{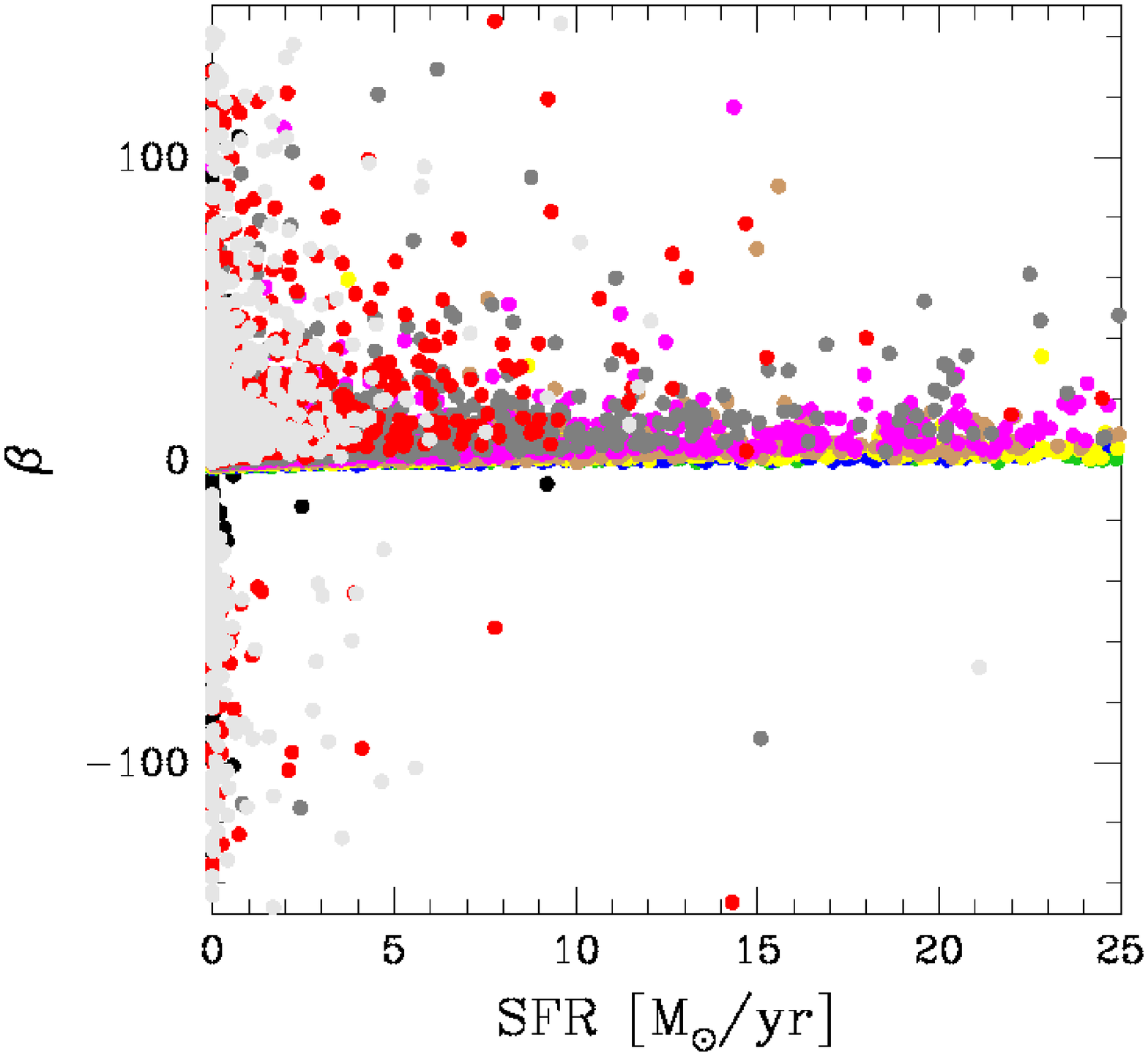}
   \caption{The $\beta-SFR$ plane. The color dots and symbols as before.}
              \label{fig:14}
    \end{figure}

\section{The important role of  $\beta$}
\label{sec:5}
To better understand the effects played by $\beta$ it is necessary to think about the possible variations of \re\ and \Ie\ when $L$ and $\sigma$ vary in the \Lsig\ plane. There are six possible changes of $L$ and $\sigma$ in this plane: $\sigma$ either decreases,  increases or remains constant, and the same does $L$. The effective relationship between the two variables depends in turn on $\beta$, e.g. when $\beta$ is negative, not necessarily there is a decrease in luminosity, and when $\beta$ is positive, a decrease in luminosity might also occurs \citep[see][for a detailed discussion of this topic]{Donofrio_Chiosi_2022,Donofrio_Chiosi_2023}. {The ambiguity in the direction of evolution can be only solved by looking at the movements of the galaxies in the different SSRs, in particular observing the behavior of \Ie.}
When the luminosity of a galaxy changes, both the effective radius $R_e$ and the mean effective surface intensity \Ie\ vary. This happens because $R_e$ is not a true  physical radius, like e.g. the virial radius (which depends only on the total mass), but it is the radius of the circle that encloses half the total luminosity of the galaxy. {Since galaxies have different stellar populations with different ages and metallicity, it is very unlikely that a change in luminosity does not change the whole shape of the luminosity profile and therefore the value of \re.} If the luminosity decreases passively, in general one could expect a decrease of $R_e$ and an increase of $I_e$. On the other hand, if a shock induced by harassment or stripping induces an increase of $L$ (and a small decrease in $\sigma$), we might expect an increase of $R_e$ and a decrease of $I_e$. 
The observed variations of these parameters depend strongly on the type of event that a galaxy is experiencing (stripping, shocks, feedback, merging, etc.).
In general, one should keep in mind that these three variables $L$, $R_e$ and $I_e$ are strongly coupled each other and that even a small variation in $L$ might result in ample changes of $R_e$ and $I_e$.

In summary, as already pointed out, the variations of the parameter $\beta$  with time are responsible of all the changes observed in the FP projections. This means that the FP problem should be considered from an evolutionary point of view where time plays an important role and the effects of evolution are visible in all the FP projections. The single SSRs are snapshots of an evolving situation. The \Lsigb\ law catches such evolution in the right way, predicting the correct direction of motion of each galaxy in the basic diagnostic planes.

We now show that the parameter $\beta$ changes with the cosmic epochs and that such variations are in turn related mainly to the change of the mean surface intensity due to the natural variation of  the star formation activity with time. We will see that $\beta$ tends to be low when star formation is high and viceversa. A large scatter is however present at all epochs.  Furthermore we will show that $\beta$ increases considerably if and when the galaxy can attain  the condition of full virialization, i.e. the two variables $M_s$ and $R_e$ combine in such a way as to yield the measured velocity dispersion (i.e. that measured for the stellar content).

Figure \ref{fig:13} shows the $\beta-\log(I_e)$ plane. The dots of different colors represent the galaxies at different redshifts using the same color code of Fig. \ref{fig:5}. From this plot it is clear that $\beta$ increases and $\log(I_e)$ on average decreases when the cosmic epoch approaches $z=0$ (light gray dots). In the remote epochs ($z=4$) and up to $z\sim1.5$ we observe an almost linear dependence of $\beta$ on \Ie, in which $\beta$ ranges from 0 to $\sim20$. This condition is an indication that at such epochs the galaxies are still far from the full virialization. The real data of WINGS (black dots) are very well superposed over the simulation data, showing a large spread with large positive and negative values of $\beta$.

This behavior of $\beta$ is connected with the average star formation rate at the different cosmic epochs. It is clearly seen in Fig. \ref{fig:14}, where we note that, when the SFR is high, the values of $\beta$ are close to 0-10. The large scatter in $\beta$ starts to be visible with the gray dots at $z\sim0.6-1.0$, the same epoch in which before we have seen the tail of {high luminosity } galaxies in the  SSRs for the first time.

\begin{figure}
   \centering
   \includegraphics[scale=0.40]{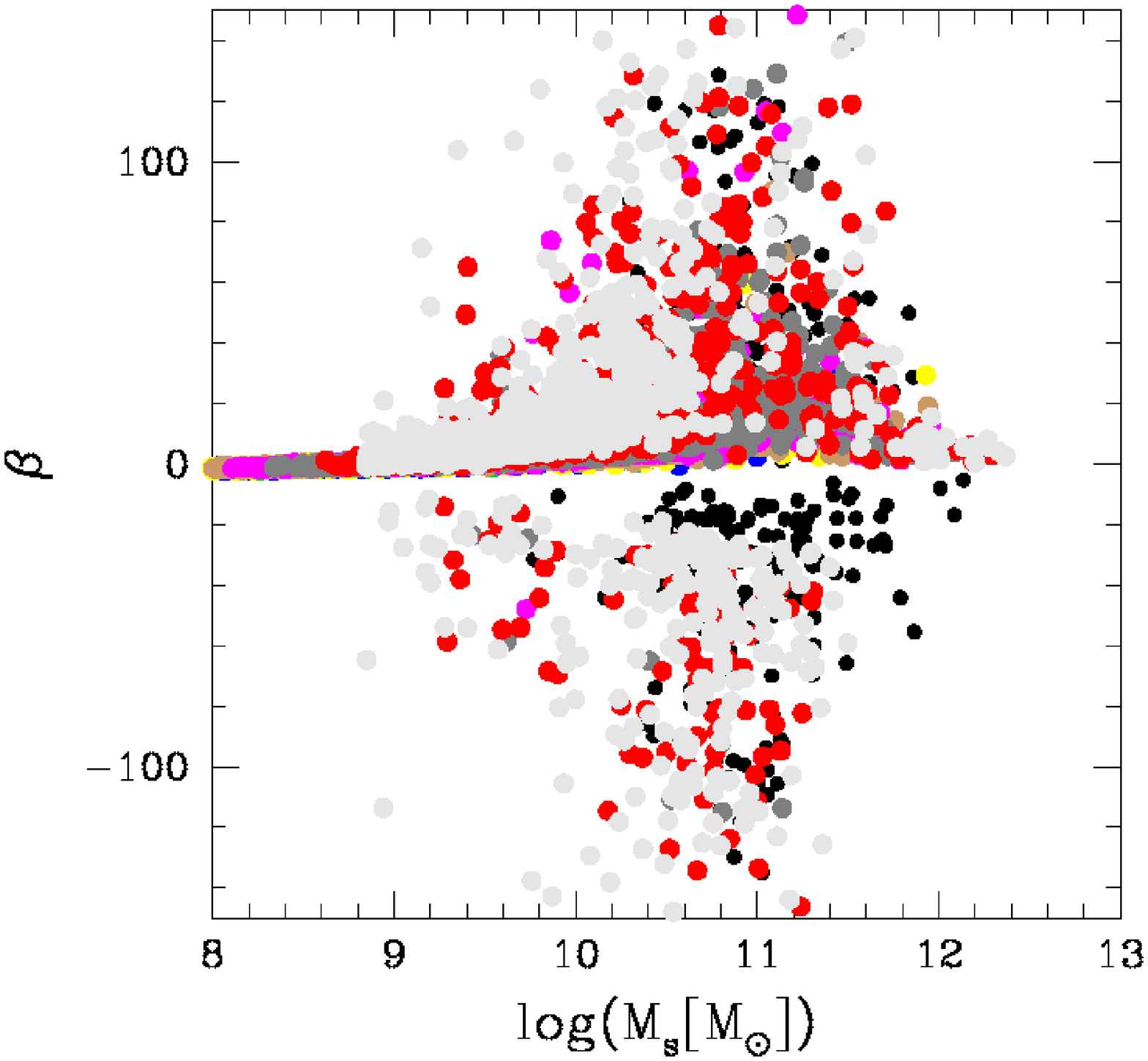}
   \caption{The $\beta-\log(M_s)$ (stellar mass) plane. The color code and symbols are the same as in previous figures.}
              \label{fig:15}
    \end{figure}

Figure \ref{fig:15}  is also helpful to understand that $\beta$ gets large positive and negative values preferably in galaxies with masses higher than $\sim 10^{10} M_\odot$.

%%%%%%%%%%%%Figure 15

%%%%%%%%%%%%Figure 16
   \begin{figure}
   \centering
   \includegraphics[scale=0.40]{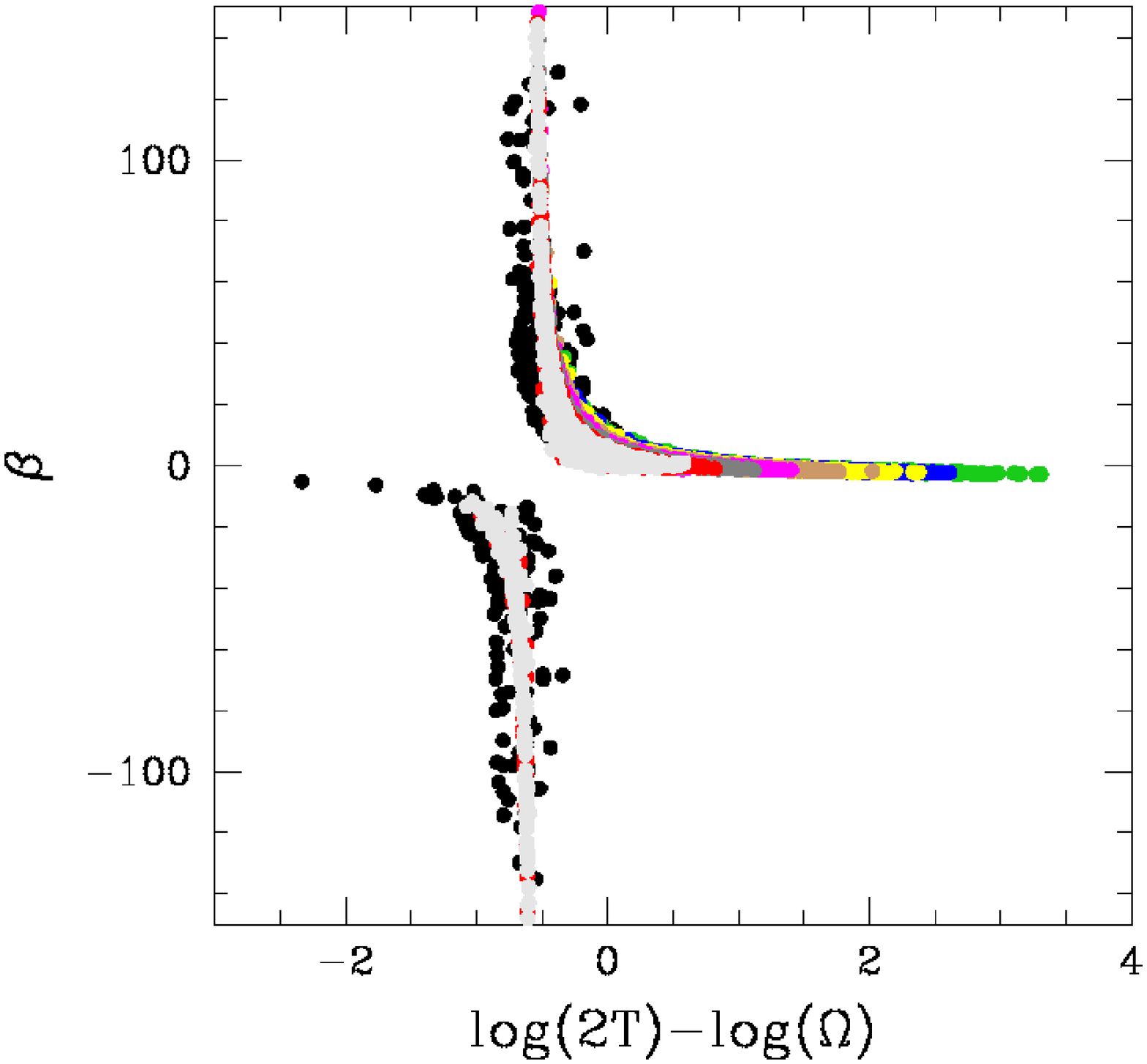}
   \caption{The $\beta-[\log(2T)-\log(\Omega)]$ plane. {  The color code and symbols are the same as in previous figures}. }
              \label{fig:16}
    \end{figure}

Finally, Fig. \ref{fig:16} shows $\beta$ versus the quantity $[\log(2T)]-\log(\Omega)]$ that is a proxy of the virial condition, being the difference between the kinetic and potential energy of the stellar systems. The figure clearly indicates that $|\beta|$ increases while $\beta$ can be either positive or negative, when the difference of the two energies approach zero\footnote{The difference predicted by the VT is 0, but the calculated energies depend on \re, which is not exactly the virial radius.}. Note that at high redshift $\beta$ remains very close to 0. This means that the galaxies are still far from  virial equilibrium. In contrast at low redshifts the peak of $\beta$ falls in the interval 0 to 20 ($z=1$) and 0 to 50 ($z=0$) with larger and larger spreads toward both high positive and low negative values. 

\begin{figure}
   \centering
 {  
   \includegraphics[scale=0.40]{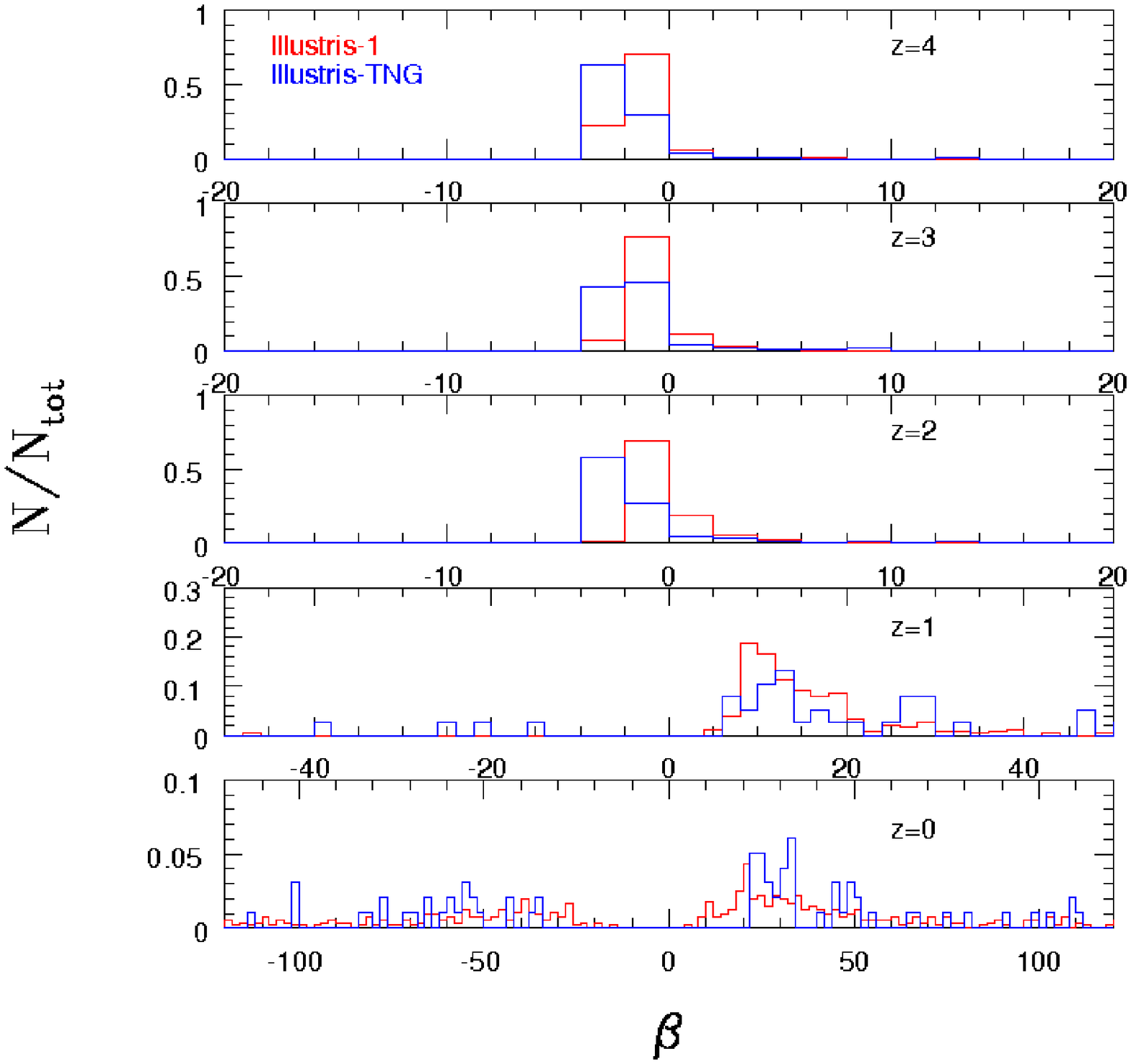}
   }
   \caption{Histogram of  $\beta$ for models galaxies of the Illustris-1 (red line) and Illustris-TNG (blue line) samples as a function of the redshift.}
              \label{fig:17}
    \end{figure}

In Fig. \ref{fig:17} we show the histogram of the number frequency distribution ($N/N_{tot}$) of $\beta$ in the model galaxies at different redshifts. There is no much difference between the histograms of the Illustris-1 and Illustris-TNG samples. The most remarkable features to note are that (i) at high redshifts { ($z \geq 2$)} the distribution peaks fall in the interval $\-4 \leq \beta \leq 0$ with a small tail of positive values in the interval $0 \leq \beta \leq 4$; (ii) the distribution  gradually spreads  to higher values of $|\beta|$ at low redshifts (1 and 0), both positive and negative values of $\beta$ are present;  (iii) finally, at low redshifts the peaks are visible both in the positive and negative range of the $\beta$ values and $|\beta|$ can attains large values.   

%%%%%%%%%%%%Figure 17 

The reason why we observe such large dispersion in $\beta$ is that the term $1- 2A'/A$ at the denominator of eq.(\ref{eqbeta}) becomes very close to zero. Consequently  both $\log(L'_0)$ and $\beta$ diverge. According to the direction in which 0 is approached, one can have either very large and positive or large and negative values for $\beta$. As already discussed, this happens when the system is in conditions of strict virialization. The sign of $\beta$ depends on the particular history of the variables $M_s$, $R_e$, $L$, and  $I_e$, in other words whether the term  $2A'/A$ is tending to 1 from below ($\beta > 0)$ or above 1 ($\beta<0$). From an operational point of view we may define "state close to strict virialization" when $|\beta| > 20$.  

Notably the Illustris-1 and Illustris-TNG  models agree very well with the observational data (black dots in these panels). The inclusion of real dynamics and the hierarchical scenario  provide much better conditions to bring the action of virialization into evidence. The hierarchical scenario by mergers, ablation of stars and gas, harassment, secondary star formation, inflation of dimension by energy injections of various kinds, etc. induces strong variations of the fundamental parameters of a galaxy and hence strong temporary deviations from the virial conditions. However, after this happened, the viral conditions  are soon recovered over a suitable timescale. This can be  short or long  depending on the amount of mass engaged in the secondary star forming activity and the amount of time elapsed since the star forming event took place \citep[see the burst experiments in][]{Chiosi_Carraro_2002, Tantalo_Chiosi_2004}. As  a consequence of all this,  detecting systems on their way back to virial equilibrium is likely a frequent event thus explaining the high dispersion seen on the $\beta$-$I_e$ plane.  The value of $\beta$ evaluated for each galaxy can provide a useful hint about the equilibrium state reached by the system. Most likely, the condition of strict virial equilibrium is a transient phenomenon that could occur several times during the life of a galaxy. This is suggested by the large number of galaxies with very small  or very high $\beta$.

%%%%%%%%%%%%Figure 18
   \begin{figure}
   \centering
   \includegraphics[scale=0.40]{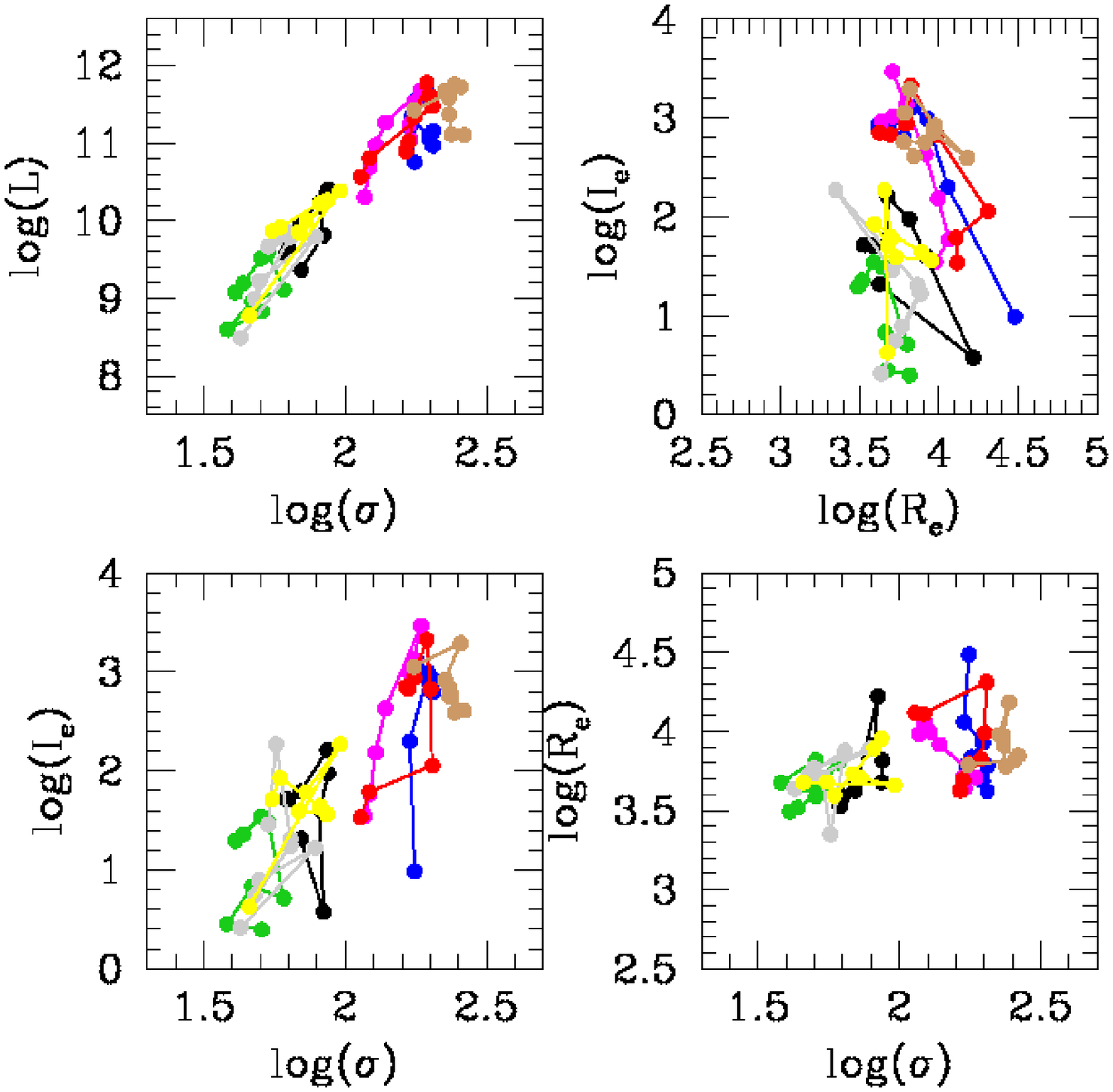}      %%%% scale=0.8
   \caption{The  SSRs for eight single galaxies. The different colors mark the single galaxies. The positions in the diagram connected by the colored lines mark the distribution at different redshift epochs.}
              \label{fig:18}
    \end{figure}

\section{The history of mass assembly}
\label{sec:6}
\begin{figure}
   \centering
   \includegraphics[scale=0.40]{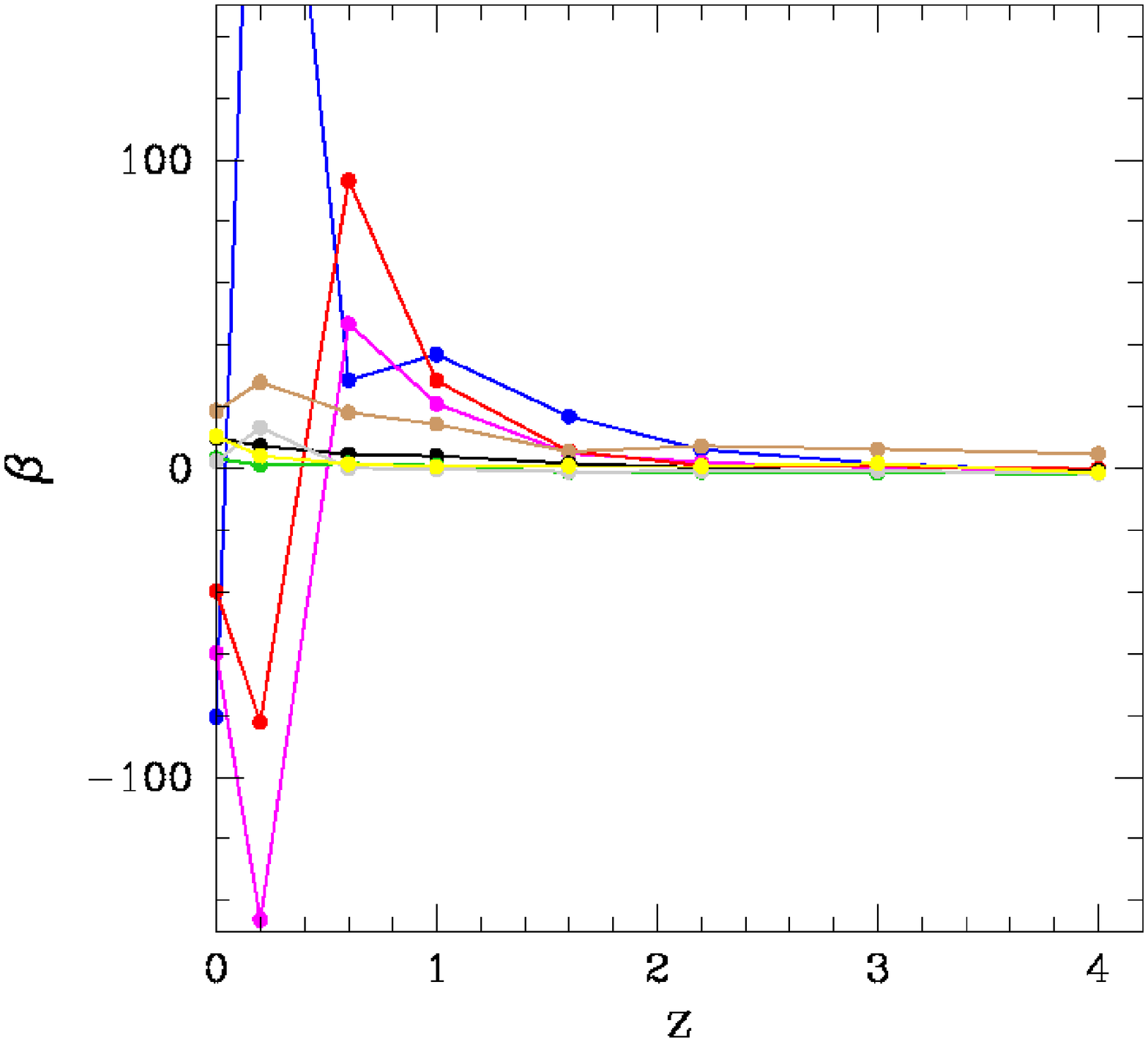}
   \caption{The $\beta-z$ plane for eight single galaxies. Colors as in Fig. \ref{fig:18}.}
              \label{fig:19}
    \end{figure}

The Illustris-1 and Illustris-TNG simulations have made clear that the history of mass assembling of galaxies is not simple, but goes through repeated episodes of mass accretion and mass removal. 
Figures \ref{fig:18} and \ref{fig:19} show the main  SSRs and the $\beta-z$ plane respectively. Eight single galaxies of different mass and evolutionary history {extracted randomly from the sample} are displayed in these plots. These  galaxies are taken from our Illustris-1 sample.  Each galaxy is indicated by a broken line of  different color, while the  mass assembly history of each object is represented by the series of dots of the same color.  Along each line there are eight points, one for each value of the redshift from $z=4$ to $z=0$ according to the list already presented in the previous sections. A very similar figure is obtained with the TNG data and therefore has not be plotted here.

The $\beta$ values of these 8 galaxies at different redshift are shown in Fig. \ref{fig:19} and are very close to 0 at every epochs with the exceptions of the blue track. Using the $\beta$ values one should enter in Table \ref{beta_values} and derive the possible directions of motion at each redshift in each of the planes.
For example the yellow and green objects have always values of beta close to 0 (a bit positive). These corresponds to slopes around $\sim-2$ in the \IeRe\ plane, $\sim2.5$ in the \Rsigma\ and $\sim-1.5$ in the \IeSig\ planes. They are therefore objects of the "big cloud" were galaxies can move in every possible direction. The blue galaxy on the other hand reaches quite high values of $\beta$ and it possible to see that its movements are in the direction close to $-1$ in the \IeRe\ plane.

It is clear from Fig. \ref{fig:18} that the galaxies do not move in the planes of the  SSRs in a continuous and uniform way, rather they randomly change their  position at different epochs. In the same figure we can also  note that the galaxies with blue, red, brown, and magenta colors are more massive, more luminous,  and with higher $\sigma$ than  the others even at early epochs ($z=4$). Even more important, we emphasizes that the same galaxies at epochs closer to us than $z\sim1.5$ are able to reach both large positive and negative values of $\beta$.  Their $\beta$s start low and gradually increase, in other words may reach the state of virial equilibrium. In contrast, the less massive and fainter galaxies have always low $\beta$s (see Fig. \ref{fig:18} and \ref{fig:19}) close to 0 and are located in the region of low $\sigma$, \Ie, and $L$. The dwarf galaxies never reach the condition of full virialization.

%%%%%%%%%%%%Figure 19

As already pointed out the condition of full virial equilibrium can be a transient state in the sense that once reached it cannot be maintained for ever if a galaxy  undergoes events such as mergers, stripping and harassment that may push it away from this condition. However, the virial equilibrium can be recovered again on a suitable time scale that of course depends on the relative intensity of the disturbing event. For instance, in the case of a merger between two galaxies of comparable mass, most likely accompanied by intense star formation, the resulting system will not be in virial equilibrium and will take quite some time to reach this condition. On the contrary, the merger between two galaxies of significantly different mass, likely accompanied by modest star formation, will only slightly depart for the virial condition. If so, we expect that after a certain  redshift,  only the massive galaxies remain unperturbed by mergers  and can move toward the condition of strict virialization, while the low mass ones are still far from this ideal {condition}. The few objects displayed in Fig. \ref{fig:19} are typical examples of the above situations. 

\section{Discussion and conclusions}\label{sec:7}

The aim of this paper is to show that combining the VT with the \Lsigb\ relation (as a proxy of evolution, in which $\beta$ and $L'_0$ vary from galaxy to galaxy and in the course of time) is rich  of positive consequences  
\citep[see][for earlier studies along this line of thought]{Donofrioetal2017,Donofrioetal2019,Donofrioetal2020,DonofrioChiosi2021}. The variation of $\beta$ and $L'_0$ with time traces the path followed by each galaxy in the various SSRs. The \Lsigb\ law together with the VT yield  \IeRe, \Rsigma, \IeSig\ and \MRa\ relations that nicely reproduce the data  and more important strongly suggest the existence of a system of two equations in the unknowns $\beta$ and $L'_0$  with coefficients functions of $M_s$, $R_e$, $L$, and $I_e$ that for each galaxy determines the value of $\beta$ and $L'_0$. With the aid of these relations we can determine the instantaneous position and direction of motion of a galaxy on the FP and its projection planes. {Because of this and limited to ETGs, we named these equations {\it fundamental equations of galaxy structure and evolution} \citep{Donofrio_Chiosi_2022,Donofrio_Chiosi_2023}. }

With this study we show that the Illustris-1 and Illustris-TNG databases give basic parameters of galaxies in satisfactory  agreement with the observational data for the galaxies at $z \simeq 0$. They indeed reproduce some distinct features observed in the FP projections, such as the tail of the bright ETGs, the ZoE, and the clumps of small mass objects. 

Basing on these simulated data, we look at the SSRs at different epochs (from redshift $z=0$ up to redshift $z=4$) highlighting their expected behaviour. In summary we show that: 

\begin{enumerate}
\item The  SSRs change with time;
    
\item The variations of the SSRs can be explained with the variation of the $\beta$ parameter driving the \Lsigbtempo\ law; 

\item When $\beta$ varies with time, a galaxy can move in the  SSRs only in some well defined directions that ultimately depend on $\beta$. These directions change with time and, going toward $z=0$, progressively acquire the slope exhibited by the most massive galaxies that lay along the tail of the bright ETGs at $z=0$;

\item The parameter $\beta$ can get both positive and negative values across time. Basing on it, we suggest that the  parameter $\beta$  can be considered as a thermometer gauging the virialization conditions. As a matter of fact $\beta$ can be  either large and positive or  large and negative when the galaxies are close to the virial equilibrium;

\item The only galaxies that can reach the  virial state are those that became  massive enough  (above $10^9-10^{10} M_\odot$) already at high redshifts ($z=4$). These are no longer disturbed by the merging events, which by the way  become rare events after $z\sim 1.5$ and/or in any case  are not influential in terms of changing the mass ratio between donor and accretor; 

\item Finally, the \Lsigb\ relation can be considered as an empirical way of catching the temporal evolution of galaxies. The values of $\beta$ (and $L'_0$) mirror the history of mass assembly and  luminosity evolution of a galaxy.
 
\end{enumerate}

The conclusion is that SSRs are full of astrophysical information about galaxy evolution.

\begin{acknowledgements}
      M.D. thanks the Department of Physics and Astronomy of the Padua University for the financial support.
\end{acknowledgements}

   \bibliographystyle{aa} % style aa.bst
   \bibliography{New_FP.bib} % your references Yourfile.bib

 \begin{appendix}
  
 \section{Mergers and Bursts of Star Formation}\label{appendix_A}
 
In order to better quantify the effect of a merger on the integrated light of a galaxy we present here an elementary model of population synthesis.  Let us consider  two galaxies with total stellar mass $M_1$ and $M_2$ and total luminosity $L_1$, and $L_2$  (either bolometric or in some passband). The luminosity is generated by  the stars already existing in each galaxy. The two galaxies are supposed to merge. The merger event may  or may be not  accompanied by star formation  induced by the merger itself.  Let $M_3$ be the mass of gas (belonging to one of the  galaxies or both) that is eventually turned in newly born stars.  For simplicity we consider this event as a unique single stellar population, SSP, of total mass $M_3$ generating a total luminosity $L_3$. If no star formation occurs at the merger $M_3=0$ and $L_3=0$. 

The total mass of the system is

$$M= M_1 + M_2 + M_3 $$ 

\noindent
and  the ratios between the mass of each component to the total mass are 

$$\alpha_1= \frac{M_1}{M} \qquad \alpha_2= \frac{M_2}{M} \qquad \alpha_3= \frac{M_3}{M}, $$ 
(with no star formation $M_3=0$). Let us suppose that the most massive object of the three is $M_1$, followed by $M_2$ and $M_3$ (with $M_2 > M_3$). Therefore, we have the sequence $\alpha_1 > \alpha_2 > \alpha_3$.  

The total luminosity is 

$$L = L_1 + L_2 + L_3$$

\noindent
and the corresponding ratios between the luminosity of each component to the total luminosity  are

$$h_1= \frac{L_1}{L}  \qquad h_2= \frac{L_2}{L} \qquad h_3= \frac{L_3}{L}.$$ 
With no star formation  $L_3=0$ and $h_3=0$. 

As the luminosity of a galaxy depends not only on the mass but also on the age (the luminosity gets fainter at increasing age) and it may undergo large and fast variations in presence of star formation, the sequence $h_1 > h_2> h_3$ can be easily violated.  It may happens that $h_2 > h_1$   and $h_3 > h_1$ and/or $h_3 > h_2$. In summary, we have defined the  following identities for the total mass and total  luminosity, in which the contribution of each component is brought to evidence

\begin{eqnarray}
  M &=& (\alpha_1  + \alpha_2  + \alpha_3) M  \nonumber \\
  L &=& ( h_1 +  h_2  +   h_3) L  \nonumber
  \label{eq_merger_burst}
\end{eqnarray}

Given these premises, we shortly present the key ingredients of our analysis, namely the relation with time of the mass and luminosity of single stars. This yields an idea of the timescale over which the mass and light of the most massive objects in any generation of stars vary with the age. Each generatios of stars forms a single stellar population (SSP) with a certain abundance of chemical elements. In a SSP stars are distributed in mass according to some initial mass function (IMF), in which typically  the number of stars per mass interval increases at decreasing stellar mass; many more faint stars of low mass than bright stars of high mass. Since stars evolve and die, the total light emitted by a SSP decreases and the SSP becomes fainter and redder with time. Finally, galaxies are made of stars born at different epochs and dying at different times. Therfore the stellar content of a galaxy can be conceived as a manifold of SSPs of different age, hence emitting different amounts of light of different colors. The total light is the integral of the light emitted by each SSP weighed on the rate of star formation over the whole history of a galaxy. All of this function of time and chemical composition. 

\begin{figure*}
   \centering
  {\includegraphics[scale=0.4]{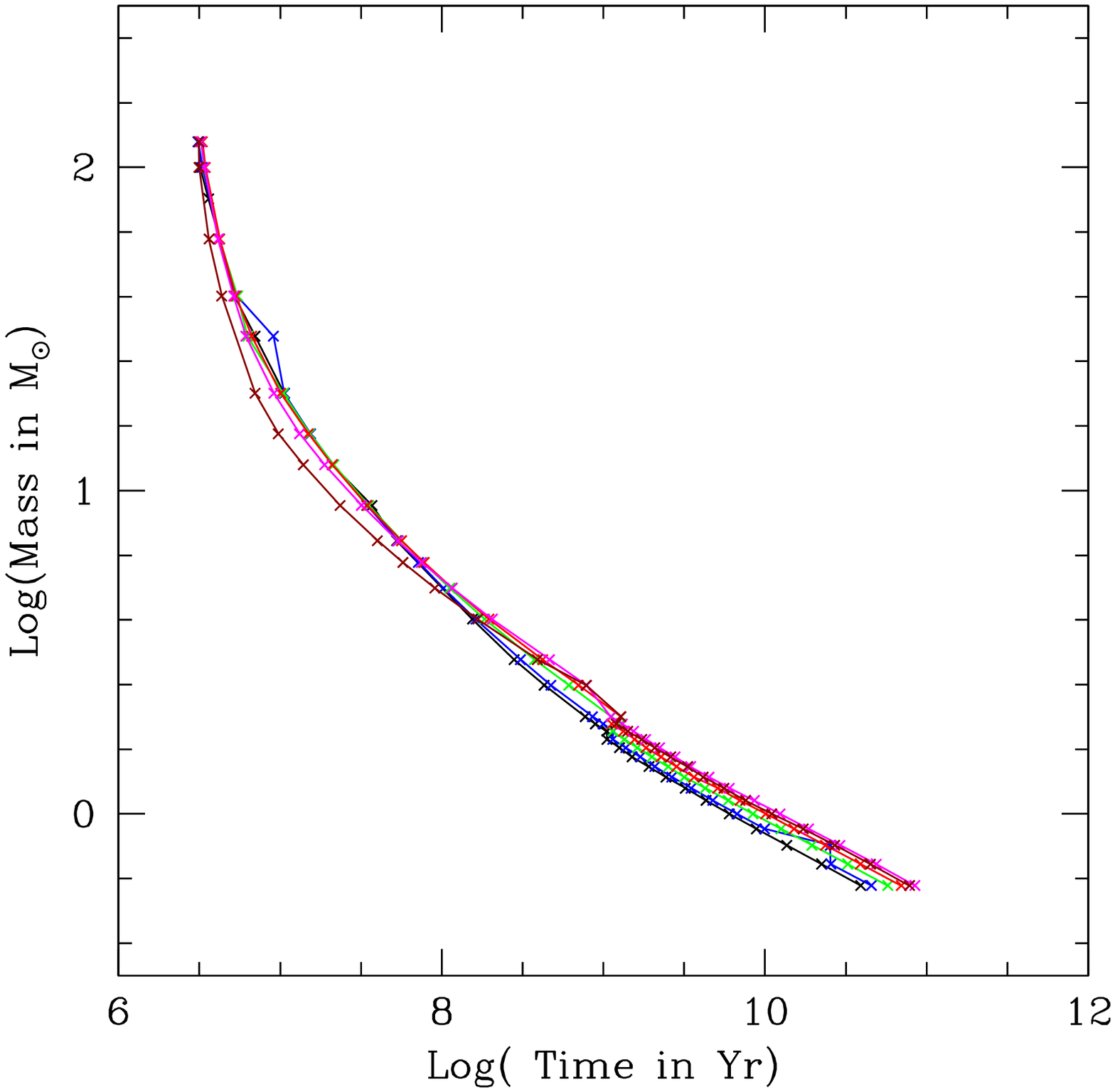}
  \includegraphics[scale=0.4]{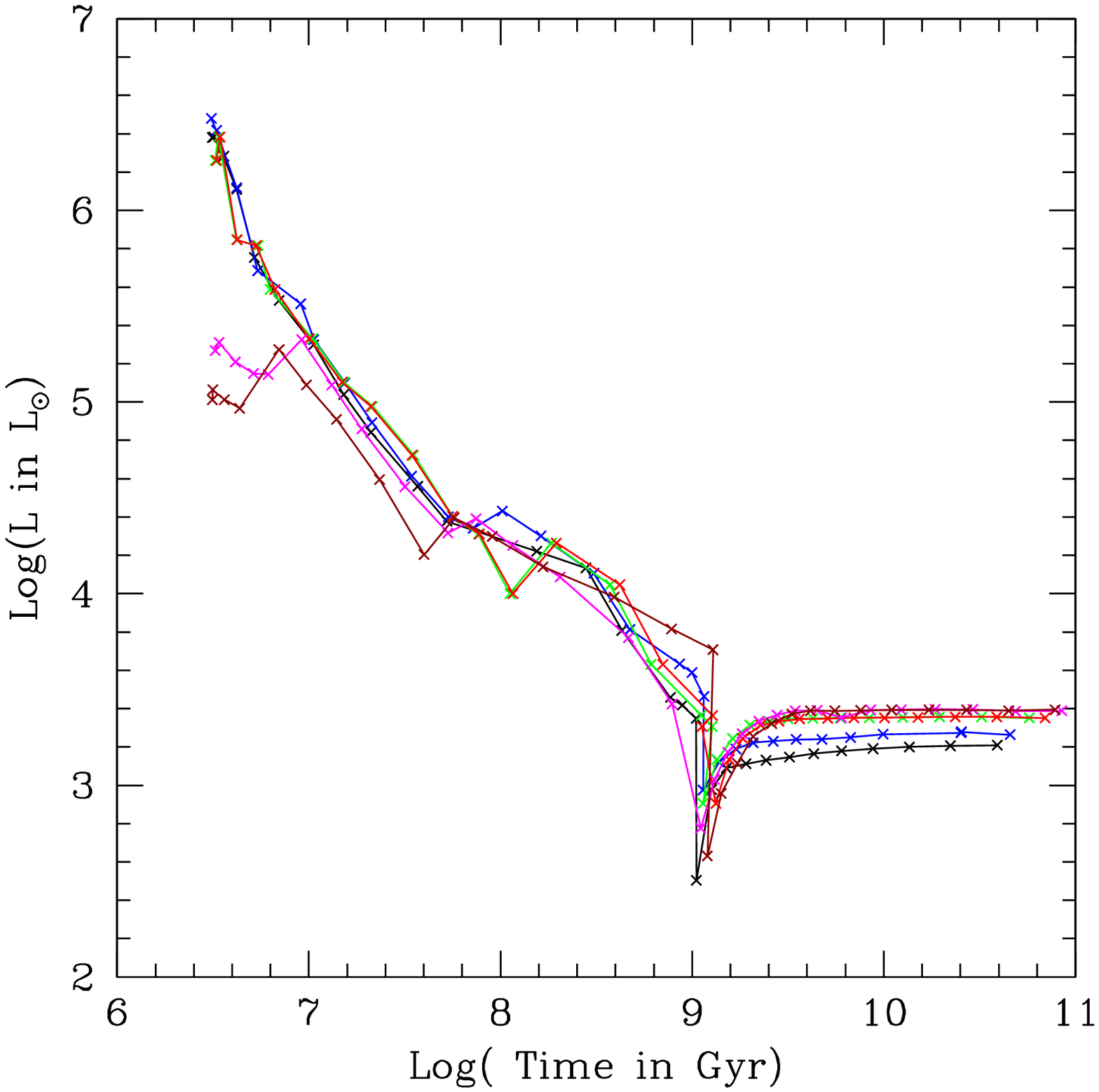} }
  \caption{
  	The stellar mass in $M_\odot$ versus age in yrs relationships (left panel). The asterisks along each curve are the star masses in the mass interval 0.6 to 120 $M_\odot$. The color code of each curve indicates the chemical composition. This labelled by the mass abundance of metals  $Z$ according to the law $\rm \Delta Y/\Delta Z=2.5$ where  Y is the mass abundance of Helium and the initial values are $\rm Z_0=0.0001$ and $\rm Y_0=0.23$. The values of the metalliciies are Z=0.0001 (black), Z=0.0004 (blue), Z=0.004 (green), Z=0.008 (magenta), Z=0.020 (red), and Z=0.050 (dark red). The mass abundance of Hydrogen is given by $\rm X=1-Y-Z$. The same as in the left panel but for the luminosity versus age relationship (right panel). The luminosity dips at ages older than 1 to 1.5 Gyrs correspond to the transition masses $\rm M_{HeF}$. Along each curve, stars more massive than $\rm M_{HeF}$ undergo quite core He-ignition, while those equal or lighter than $\rm M_{HeF}$ undergo core He-flash. }
  \label{fig:mass_lum_time}
  \end{figure*}

\begin{figure*}
  \centering 
  {   \includegraphics[scale=0.4]{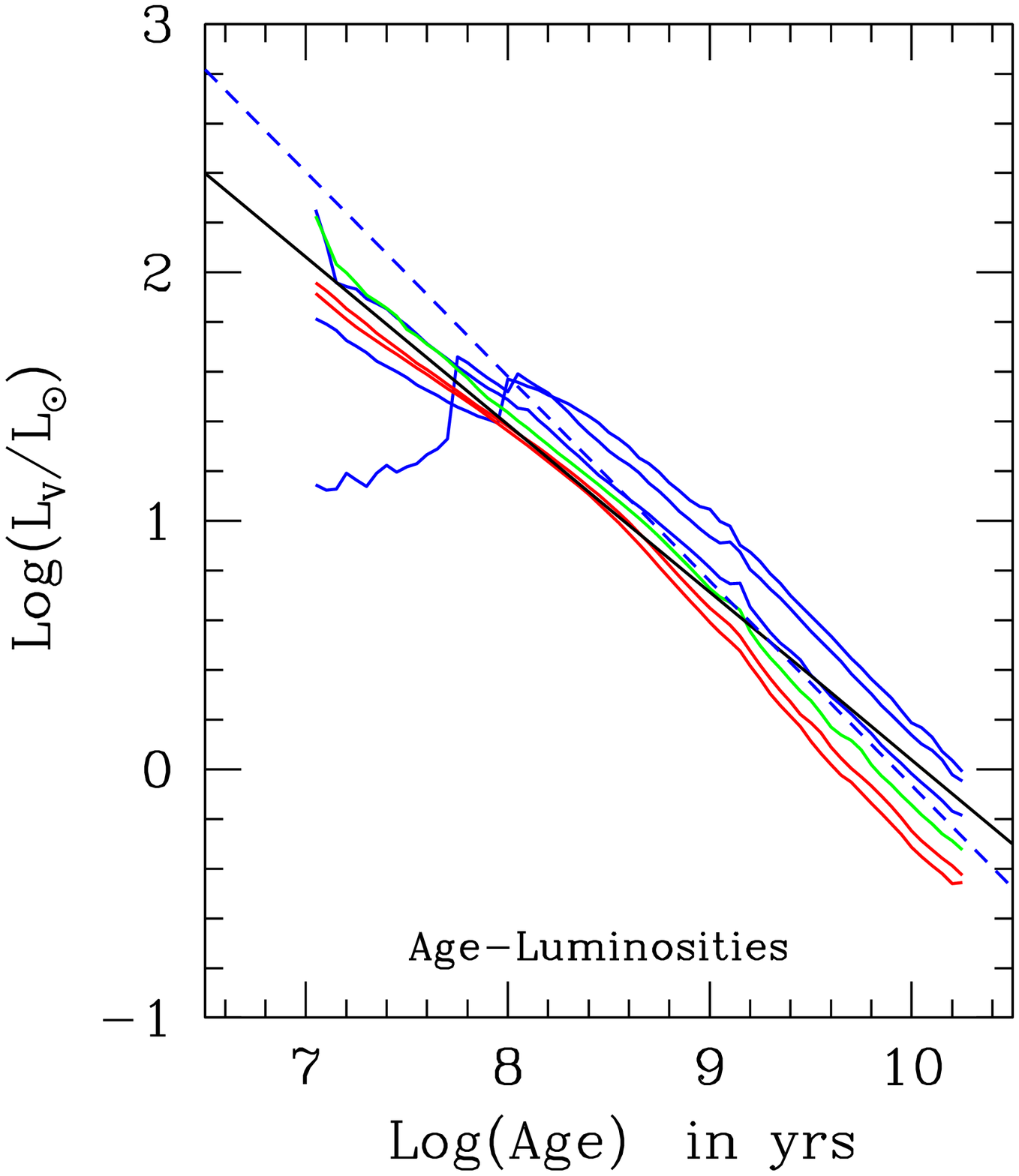}
   \includegraphics[scale=0.4]{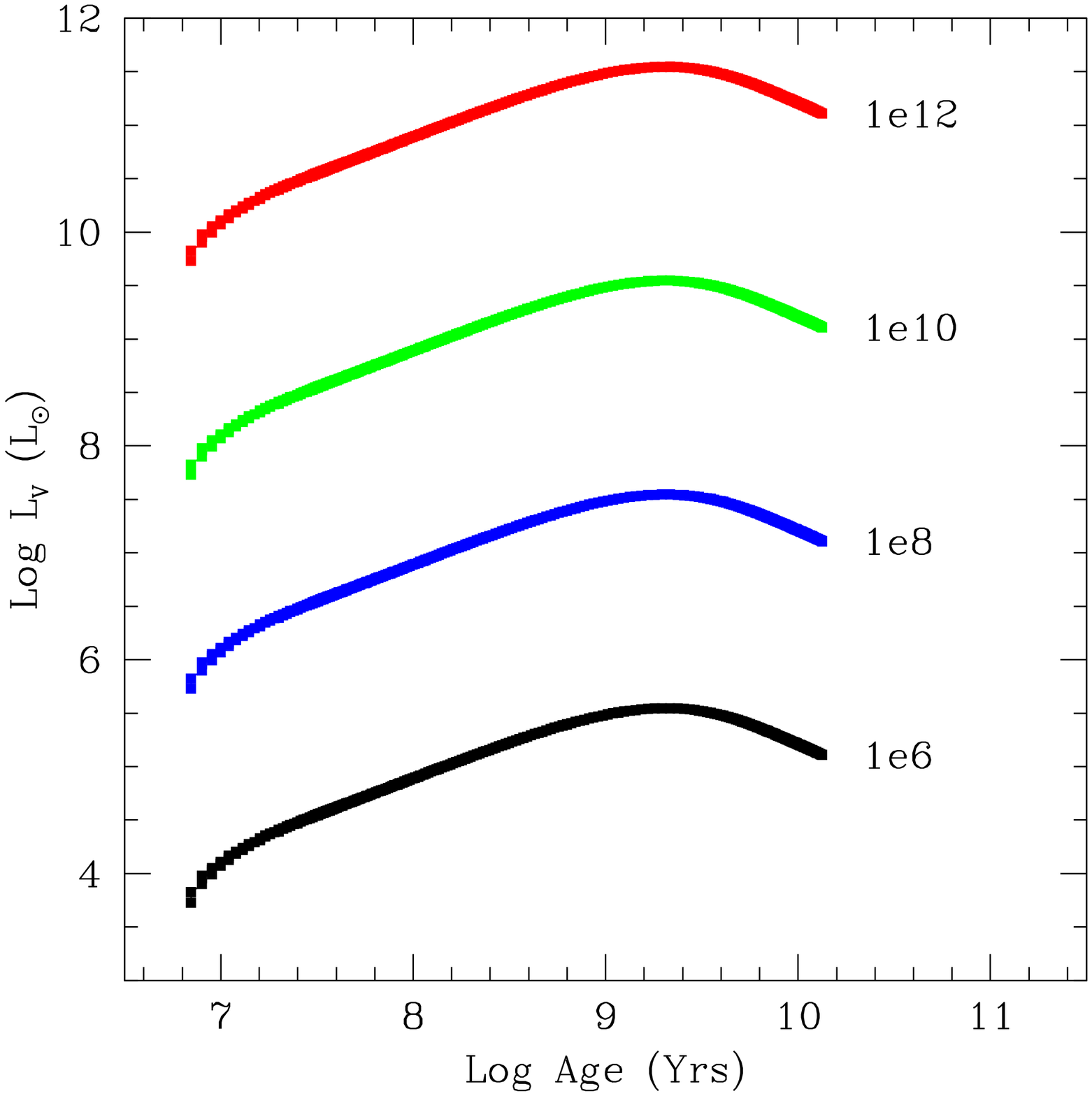} }
   \caption{The luminosity in $L_\odot$ versus time in yrs of Single Stellar Populations (SSPs) (left panel) with the Salpeter Initial Mass Function (in number of stars per mass interval) with exponent -2.5 and total SSP mass of 5.826 $M_\odot$. The chemical compositions of the SSPs and associated color code are the same as in Fig. \ref{fig:mass_lum_time}. Model galaxies with infall calculated by the authors (righ panel). The color code indicates the total baryonic mass of the galaxy in $M_\odot$ whose value is also displayed  in the label of each curve. The time scale of gas infall is 
   $\tau = 1$ Gyr. The peak in luminosity corresponds to the maximum star formation rate. The galaxy models are described by \citet{Donofrio_Chiosi_2023}.  }
   \label{fig:lum_ssp_gal_time}
    \end{figure*}

In Fig. \ref{fig:mass_lum_time}  we show the time dependence of the mass (left panel) and luminosity (right panel) of single stars (lifetime and luminosity are taken at the stage of  the brightest luminosity attained by the long lived evolutionary stages) of different mass (in the interval 0.6 to 120 $M_\odot$). In both panels the colors indicate the metallicity and the ticks along each curve mark the   value of the mass; blue is for low metal abundance and red for high metal contents.  In Fig. \ref{fig:lum_ssp_gal_time}  we show the luminosity versus time relationship for SSPs of different chemical composition as indicated (left panel), and model galaxies of different mass as indicated (right panel). In the left panel we display the luminosity (in solar units) in the V-passband of the Johnson-Bessell system and  of the SSPs with different metallicity (solid  lines with different colors where blue  is for low metal content and red for high metal content). The black solid line is the best fit of the luminosities for metallicities. In the same panel and limited to the bestfit we also show the luminosity in the B-passband (dashed line). In the right  panel we display the V-luminosity in the same photometric system of model galaxies with infall during the whole lifetime of the galaxies. This is the luminosity integrated over the many generations of stars formed in the galaxy under a suitable star formation rate. Chemical enrichment of the gas out which stars are formed is taken into account.       
 Stellar models and  SSPs, and model galaxies are taken from the Padua library of stellar models, isochrones, and SSPs  \citep[see][for all details and ample referencing]{Bertelli_etal_1994,Girardi_etal_1996,Bertelli_etal_2008,Nasi_etal_2008}. The model galaxies are calculated by the authors and are described in \citet{Donofrio_Chiosi_2023}. The SSPs in use here are for the Salpeter initial mass function with slope x=2.35 (in number of stars per mass interval), the SSP mass and luminosity are named $M_{SSP}$ and $L_{SSP}$ with  $M_{SSP}=5.826 M_\odot$.

%%%%%%%%%%%%Figure 12bis
  
  %%%%%%%%%%%%%%%%%%Figure 12 tris

Notable features of these diagrams are: 

(i) The luminosity of single stars may vary by more than two orders of magnitude and the lifetime goes from a few million years (Myrs) to more that ten billions years (Gyrs) as the mass decreases from to 120 to 0.6 $M_\odot$ with little dependence on the initial chemical composition (at least for our purposes). The interval of variation is much narrower passing to SSPs and even more to galaxy models. In SSPs, this is simply due to the integration over the mass under some initial mass function; the more numerous low mas stars of lower luminosity somewhat quench the luminosity of the brighter less numerous stars. In galaxy models, supposing that each star formation event in the time interval $dt$ can be represented by a SSP of suitable chemical composition emitting the total luminosity $l_{ssp}(t)$, the total light $L$ is given by 

$$ L = \int_0^{T_G} \Psi(t) l_{ssp}(t) dt $$
where $\Psi(t)$ is the current rate of star formation in suitable units.

(ii) The remarkable reduction in the luminosity interval passing from single stars to SSPs and model galaxies is simply due to  the integration of the contribution of single stars to the total light of the SSPs of given age  and the integration over time of the contributions from the many generations of SSPs  weighed  on the star formation rate.

(iii) The luminosity of SSPs varies by about a factor of hundred with little dependence on the chemical composition passing from young to old SSPs. From these SSPs we can derive the following bestfits
\begin{eqnarray}
 \log l_V &=& -0.685 \log T +6.886   \nonumber \\
 \log l_B &=& -0.835 \log T +8.278   \nonumber
\end{eqnarray}

\noindent
where $l_V$ and $l_B$ are the luminosities in solar units in the V and B passdands of the Johnson-Bessell photometric system  and the  age $T$ is in years. These luminosities are for the SSP mass of $M_{SSP}=5.826 M_\odot$. To be applied to a galaxy of stellar mass $M_s$ we must apply the transformations

\begin{eqnarray}
 \log L_V &=&  \log l_V - \log M_{SSP} + \log M_s   \nonumber \\
 \log L_B &=&  \log l_B - \log M_{SSP} + \log M_s   \nonumber
\end{eqnarray}
where $M_s$ is in solar units and it is identified with the galaxy mass.

(iv) In the  model  galaxies, initially the luminosity increases by a factor of ten up to a peak corresponding to the maximum of the star formation rate at about 1.5 Gyr and then decreases by about the same factor down to the present day value. 

(v) As expected the total luminosity varies by orders of magnitude among galaxies of different mass.

All this has a immediate effect on the maximum variations of the total luminosity in mergers among galaxies of different mass and intensity of the companion star formation during the merger event.  To illustrate the point we present here a few examples corresponding to typical situations occurring among real galaxies:  

\textsf{ Dry mergers (no additional star formation)}.  With this assumption we have always   $M_3=0$ and $L_3=0$. 

(a) The case of two identical objects (same mass,  same stellar content, and same age) with e.g. $M_1 = M_2 = 10^{12} M_\odot$ and luminosity $L_1 = L_2 = 10^{11} L_\odot$. The luminosities have been derived from the models in Fig.\ref{fig:lum_ssp_gal_time} (right panel) at the present age. In this case  the mass and luminosity of the resulting object are two times the starting value ($\alpha_1=\alpha_2 =0.5$, $h_1=h_2=0.5$).  

(b) The case of two objects with different mass but same age (old). Suppose $M_1 = 10^{12} M_\odot $, 
$L_1= 10^{11} L_\odot$, and  $M_2 = 10^{11} M_\odot $,  $L_2= 10^{10} L_\odot$.  The total mass is 
$M= 1.1\times 10^{12} M_\odot$ and the total luminosity is  $L= 1.1\times 10^{11} L_\odot$. Therefore,
$\alpha_1=0.909$, $h_1=0.909$  and  $\alpha_2=0.091$, $h_2=0.091$. 
The total luminosity is $L = (0.909  + 0.091) L$, i.e. the light is  dominated by the  more massive galaxy.  The merger has increased the total luminosity by about 10\%. This is a sort of lower limit because a dry merger between two galaxies of the same age differing in mass by more that a factor of ten in practice would be undetectable. 

(c) In the case of two galaxies with different mass and different age, hence different luminosities, the total luminosity may show  the effect of the younger object.   Consider the following galaxies $M_1 = 10^{12} M_\odot $, $L_1= 10^{11} L_\odot$ (the old object) and  $M_2 = 10^{11} M_\odot $, 
$L_2= 5\times 10^{10} L_\odot$ (the young object, taken near the peak value). After merging, the total mass is $M= 1.1 \times 10^{12} M_\odot$ while the total luminosity is $L \simeq 1.5\times10^{11} L_\odot$. Therefore,
$\alpha_1=0.909$, $h_1=0.667$  and  $\alpha_2=0.091$, $h_2=0.333$.  The contribution from the less massive but younger object is important, about half  that produced by  the more  massive but older object. 
Other combinations of masses and ages hence luminosities  can be easily derived from eq.(\ref{eq_merger_burst}).
Effects due to differences in the mean chemical composition of the stellar contents can be ignored at a first order approximation. 

\textsf{Wet mergers (with additional star formation)}. In this case  in  eq.(\ref{eq_merger_burst}) $M_3 \neq 0$ and $L_3 \neq 0$. The major difference with respect to previous cases is that the stellar activity induced by the merger can be approximated by a single giant SSP with its own mass and luminosity. The total light emitted by $M_3$   is $L_3 = (L_{SSP}/M_{SSP})\times M_3$. Now the interval spanned by the luminosity at aging SSP is much wider than before so depending on the age the effects can be  large. For the merging galaxies we assume  $M_1 = 10^{12} M_\odot$, $L_1=10^{11} L_\odot$, and  $M_2=  10^{11} M_\odot$, $L_2=  10^{10} L_\odot$, finally for the mass of the giant SSP simulating the induced star formation event we adopt 
$M_3= 10^{10} M_\odot $. The total mass is $M=1.11 \times 10^{12} M_\odot$. For the associated luminosity we adopt three values corresponding to a very young age, high luminosity SSP with $L_3= 1\times 10^{11} L_\odot$; an intermediate age, lower luminosity SSP with $L_3= 1\times 10^{10} L_\odot$;  and an old age, low luminosity SSP with $L_3= 1\times 10^{9} L_\odot$. In the first case the total luminosity is $L= 2.1\times 10^{11} L_\odot$, whose relative components are $h_1=0.476$, $h_2= 0.048$, and $h_3=0.476$. The burst of star formation contributes to half of the total light and 1\% of the total mass. This is a rapidly transient situation that on a short time scale fades down. In the second case, the total light is $L=1.2\times 10^{10} L_\odot$, and the relative contributions to the luminosity are $h_1= 0.834$, $h_2=0.083$, $h_3=0.083$; the burst of star formation contributes to about 10\% of the light equal to the contribution of the less massive, old galaxy. In the last case, the SSP is very old, the three relative contributions are $h_1=0.908$, $h_2=0.091$, and $h_3=0.009$. The occurrence of the burst is nearly undetectable.  
  Other combinations of the parameters can be tested with similar results.  
  
What we learn from these simple tests is that in the case of dry mergers but for the merger between objects of similar mass in which the mass and light are increased by a factor of two, the effect of mergers among objects with  different masses and ages scarcely affect the light of the originally dominant object. Wet mergers with induced star formation more efficiently leave their fingerprints on the post-merger light of a galaxy. Unfortunately the bright phase is of short duration, the time scale required by the SSP to evolve from a turnoff, in the range of bright massive stars (say 20 $M_\odot$ with a lifetime of a few Myr),  down to a turnoff in the range of low mass stars (say below 2 $M_\odot$ with a lifetime of about 1 Gyr), see the panel of Fig. \ref{fig:mass_lum_time}. Therefore,  it is by far more probable to catch a galaxy that underwent a wet merge when the burst of star formation has already faded down  to low luminosities.   
This is an interesting result that could explain why the Faber-Jackson relation and the Fundamental Plane we see today show little scatter in the observational distribution of galaxies.

 \end{appendix}

\end{document}